\documentclass[aps,prl,epsfigure,twocolumn,superscriptaddress]{revtex4-1}
\usepackage[colorlinks=true,linkcolor=blue,urlcolor=blue,citecolor=blue,pdfusetitle]{hyperref}
\usepackage[utf8]{inputenc}
\usepackage[english]{babel}
\usepackage{amsmath}
\usepackage[caption = false]{subfig}
\usepackage{graphicx,epstopdf}
\usepackage{blindtext}
\usepackage[table,xcdraw]{xcolor}
\usepackage{lipsum}
\usepackage{amsfonts}
\usepackage{bbm}
\usepackage{amssymb}
\usepackage{enumerate}
\usepackage{color}
\usepackage{latexsym}
\usepackage{physics}
\usepackage{times,txfonts}

\newcommand{\Ccal}{\mathcal{C}}
\newcommand{\Dcal}{\mathcal{D}}
\newcommand{\Ecal}{\mathcal{E}}

\newcommand{\Lcal}{\mathcal{L}}

\newcommand{\Ncal}{\mathcal{N}}

\newcommand{\1}{\mathbbm{1}}

\newcommand{\sket}[1]{| #1 \rangle}

\usepackage{orcidlink}

\newcommand{\UFSCar}{Departamento de Física, Universidade Federal de São Carlos, \\Rodovia Washington Luís, km 235 - SP-310, 13565-905 São Carlos, SP, Brazil}

\newcommand{\SU}{Department of Physics, Stockholm University, AlbaNova University Center 106 91 Stockholm, Sweden}

\newcommand{\Nice}{Universit\'e C\^ote d'Azur, CNRS, Institut de Physique de Nice, 06560 Valbonne, France}

\begin{document}

	\title{Generation of maximally-entangled long-lived states with giant atoms in a waveguide}

\author{Alan C. Santos}
\email{ac\_santos@df.ufscar.br}
\affiliation{\UFSCar}
\affiliation{\SU}

\author{R. Bachelard}
\email{romain@ufscar.br}
\affiliation{\UFSCar}
\affiliation{\Nice}


\begin{abstract}
	In this paper we show how to generate efficiently entanglement between two artificial giant atoms with photon-mediated interactions in a waveguide. Taking advantage of the adjustable decay processes of giant atoms into the waveguide, and of the interference processes, spontaneous sudden birth of entanglement can be strongly enhanced with giant atoms. Highly entangled states can also be generated in the steady-state regime when the system is driven by a resonant classical field. We show that the statistics of the light emitted by the system can be used as a witness of the presence of entanglement in the system, since giant photon bunching is observed close to the regime of maximal entanglement. Given the degree of quantum correlations incoherently generated in this system, our results open a broad avenue for the generation of quantum correlations and manipulation of photon statistics in systems of giant atoms.
\end{abstract}

\maketitle

Driven artificial atoms in a waveguide constitute the heart of superconducting quantum circuits~\cite{wallraff2004}, in which the photon-mediated interaction between distant atoms can lead to collective effects such as sub-radiance/superradiance~\cite{Arjan:13,Mlynek:14}, and allows for the creation and transport of strongly correlated photons~\cite{Shen:07,Zheng:12}. Such systems have been used to efficiently create maximally entangled states of $N$ two-level atoms~\cite{Wang:20}, which are important resources for quantum information processing~\cite{Gottesman:99,dicarlo2010} and quantum communication~\cite{Bennett:93,Ren:17}, for example. However, maintaining this entanglement over time is all the more challenging because of decoherence processes. This has lead to the development of strategies to harness this decoherence and generate entanglement between sub-systems, for instance, by coupling the system whether to thermal baths of negative temperature~\cite{Tacchino:18} or fermionic reservoirs~\cite{Wang:PRA19}, by exploiting memory effects of the environment~\cite{Huelga:12}, or through the coherent control of non-degenerate atoms in cavities~\cite{Oliveira:22}.

In this scenario, identifying configurations where the effect of the environment on the entanglement is reduced to its minimum is an important task for the development of quantum technologies. In the frame of waveguide quantum electrodynamics, a promising alternative to small emitters has been introduced: \textit{giant atoms} are emitters whose coupling with a guided wave extends over such a length that their interaction cannot be described as that of point-like emitters~\cite{Kockum:14,Kockum:20}. In particular, the dipole approximation does not hold any longer~\cite{Manenti:17,Noguchi:17,Bolgar:18}. The peculiar characteristics of giant atoms, such as frequency-dependent relaxation rates~\cite{Kockum:14,Vadiraj:21,Wang:21,Du:22,Zhao:20} and decoherence-free mediated interaction between two atoms~\cite{Kockum:18,Soro:22}, make this system an ideal platform for the robust generation of entanglement~\cite{Kannan:20,Yu:21}. The multiple connections between each atom and the waveguide, and the possibility to intertwine the different atoms through their connection points, opens a broad field of perspectives for the manipulation of entangled states and for information processing, which has barely been scratched up to now~\cite{Kockum:20}.

\begin{figure}[t!]
	\centering
	\includegraphics[width=\columnwidth]{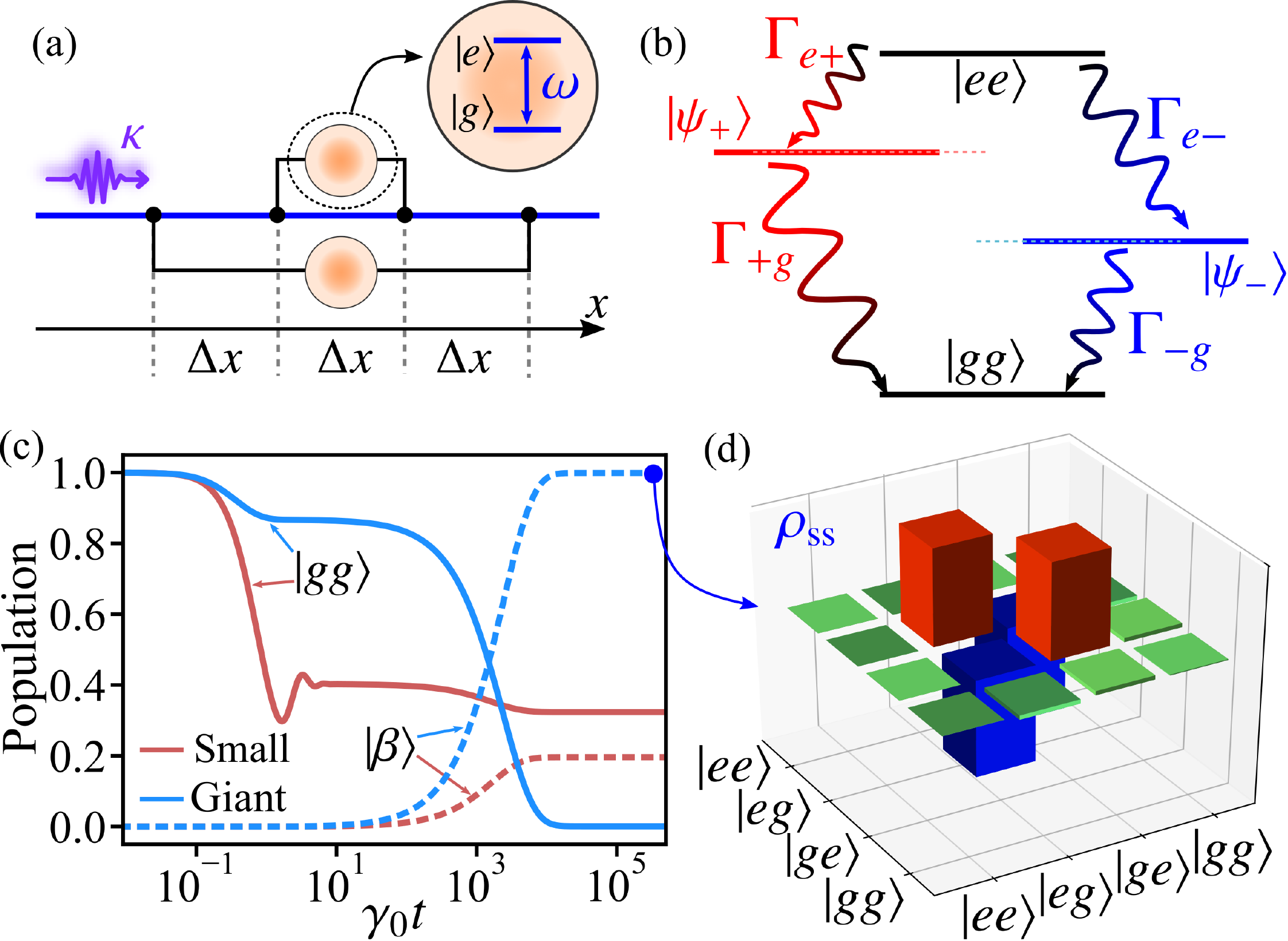}
	\caption{(a) Schematic representation of the two giant atoms (with ground and excited states $\ket{g}$ and $\ket{e}$, respectively) in the ``nested'' configuration considered in the present work. (b) The energy diagram shows the collective states of the system, and the decay rates between them. (c) Population dynamics of the ground state $\ket{gg}$ and the (maximally entangled) Bell state $\ket{\beta}\!=\!(\ket{ge}-\ket{eg})/\sqrt{2}$, for the nested giant atoms with $\kappa\Delta x/\pi\!=\!0.01$ and Rabi frequency of the external field $\Omega_{0}\!=\!1.5\gamma_{0}$. (d) State tomography of the steady-state density matrix $\rho_{\text{ss}}$ of the system, which corresponds to $\rho_{\text{ss}}\!\approx\!\ket{\beta}\bra{\beta}$ obtained in the dynamics considered in (c).}
	\label{Fig1} 
\end{figure}	

In this paper we study the generation of entanglement for two giant atoms through non-unitary (incoherent) decay processes. More specifically, we show how two giant atoms within a tunable frequency $\omega_{0}$ of the light in the waveguide, see sketch in Fig.~\ref{Fig1}{\color{blue}a}, can be used to create, control and engineer metastable entangled state. Adjusting the frequency $\omega_{0}$, we are able to tune the asymmetry of the decay rates in the different decay channels, as shown in Fig.~\ref{Fig1}{\color{blue}b}. We take advantage of this property to stimulate the sudden birth of concurrence, generating quantum correlations one order of magnitude larger than for small atoms. In addition, adjusting the external control field and the frequency $\omega_{0}$, we show how long-lived entangled states can be efficiently created through a strategy of population inversion in the steady state  (Figs.~\ref{Fig1}{\color{blue}c} and~\ref{Fig1}{\color{blue}d}). Differently from previous studies, our scheme does not require reservoir engineering~\cite{Tacchino:18,Wang:PRA19}, non-Markovian effects~\cite{Huelga:12} or energy level engineering~\cite{Oliveira:22}: the unique properties of giant atoms in terms of decoherence and energy levels are exploited to prepare maximally-entangled steady state. Finally, we show that the statistics of the light emitted by the two giant atoms in the waveguide can be used as a entanglement witness in the system, so the entanglement can be monitored in a non-destructive way through the scattered light.

\emph{Coupled dynamics of giant atoms --} Let us consider $N$ giant two-level atoms interacting with a one-dimensional waveguide. Atom $k$ is coupled to the waveguide in $K_k$ points, at positions $\{x^{(k)}_{n}\}$ and with an associated decay rate $\gamma^{(k)}_{n}$ (with $k=\{1,\cdots,N\}$ and $n=\{1,\cdots,K_{k}\}$). In the weak coupling limit and using the rotating wave approximation, the effective dynamics of the atomic system is described by a density matrix $\hat\rho$ obtained by tracing out the bosonic modes of the waveguide~\cite{Kockum:14,Kockum:18}:
\begin{align}
	\frac{d}{dt} \hat\rho(t) = \frac{1}{i\hbar} [\hat H_{0} + \hat H_{\text{cc}},\hat\rho(t)] + \Lcal[\hat\rho(t)] , \label{Eq-MasterEqGiant}
\end{align}
with $\hat H_{0}=\sum_{n=1}^{N} \hbar (\omega_{n} + \delta_{n})\hat\sigma^{+}_{n}\hat\sigma^{-}_{n}$ the Hamiltonian of the independent atoms. The frequency shift $\delta_{k}=(1/2)\sum_{\ell = 1}^{K_{k}} \sum_{m = 1}^{K_{k}} (\gamma^{(k)}_{\ell}\gamma^{(k)}_{m})^{1/2} \sin \varphi^{k,k}_{\ell,m}$ of each atom stems from the phase $\varphi^{k,n}_{\ell,m}=\kappa |x^{(k)}_{\ell} - x^{(n)}_{m}|$ acquired by the wave as it propagates between the connection points $x^{(k)}_{\ell}$ and $x^{(n)}_{m}$. With $\kappa$ the wavenumber of the light in the waveguide. The coupling between the atoms is composed of a coherent excitation-exchange component $\hat H_{\text{cc}}=\hbar \sum^{N}_{j\neq n} \sum^{N}_{n=1} \Delta_{jn} \hat \sigma^{+}_{j}\hat \sigma^{-}_{n}$, with  $\Delta_{jn}=(1/2)\sum^{K_{j}}_{\ell}\sum^{K_{n}}_{m} (\gamma^{(j)}_{\ell}\gamma^{(n)}_{m})^{1/2} \sin\varphi^{j,n}_{\ell,m}$, and a dissipative part given by the following Lindbladian:
\begin{align}
	\Lcal [\hat\rho(t)] = \frac{1}{2} \sum_{j,n=1}^{N,N} \Gamma _{jn} \left[ 2 \hat\sigma^{-}_{j}\hat\rho(t)\hat\sigma^{+}_{n} - \{\hat\sigma^{+}_{j}\hat\sigma^{-}_{n},\hat\rho(t)\} \right] ,
\end{align}
with $\Gamma_{jn}=\sum^{K_{j}}_{\ell}\sum^{K_{n}}_{m} (\gamma^{(j)}_{\ell}\gamma^{(n)}_{m})^{1/2} \cos\varphi^{j,n}_{\ell,m}$ the cross decay term. The terms $\Gamma_{jj}=\Gamma_{j}$ corresponds to the single-atom decay rate, that is, in absence of the other giant atoms. Differently from small atoms, this rate depends on the relative phase between the connection points of the given atom, and it is not simply the sum of the decay rate at each connection point of that atom. In addition, the atoms are driven by an external field, which adds the extra Hamiltonian term (in the rotating frame): $\hat{H}_{\text{ext}}\!=\! \hbar\sum_{n=1}^{2} \Omega_{0}( \hat{\sigma}_{n}^{+} + \hat{\sigma}_{n}^{-})$, with $\Omega_{0}$ the Rabi frequency of this pump. In superconducting qubits, it can be done with independent drive lines applied to the atoms~\cite{Barends:14,krantz:19}, for instance, in which case the Lindblad form of the Eq.~\eqref{Eq-MasterEqGiant} does not change~\cite{SupInf_GiantAtoms}. 
Note that the above equations are $\kappa$-periodic, meaning that retardation effects associated with the propagation time of photons in the waveguide are not accounted for. Non-Markovian dynamics~\cite{Lei:22,Wang:22,Lei:22b} can lead to disentanglement and entanglement revival~\cite{Qiu:22,Yin:23}, yet only when the propagation time becomes $t\sim 1/\gamma_0$. In recent experiments~\cite{Kannan:20}, this corresponds to meter-size waveguides.

\begin{figure}[t!]
	\centering
	\includegraphics[width=\columnwidth]{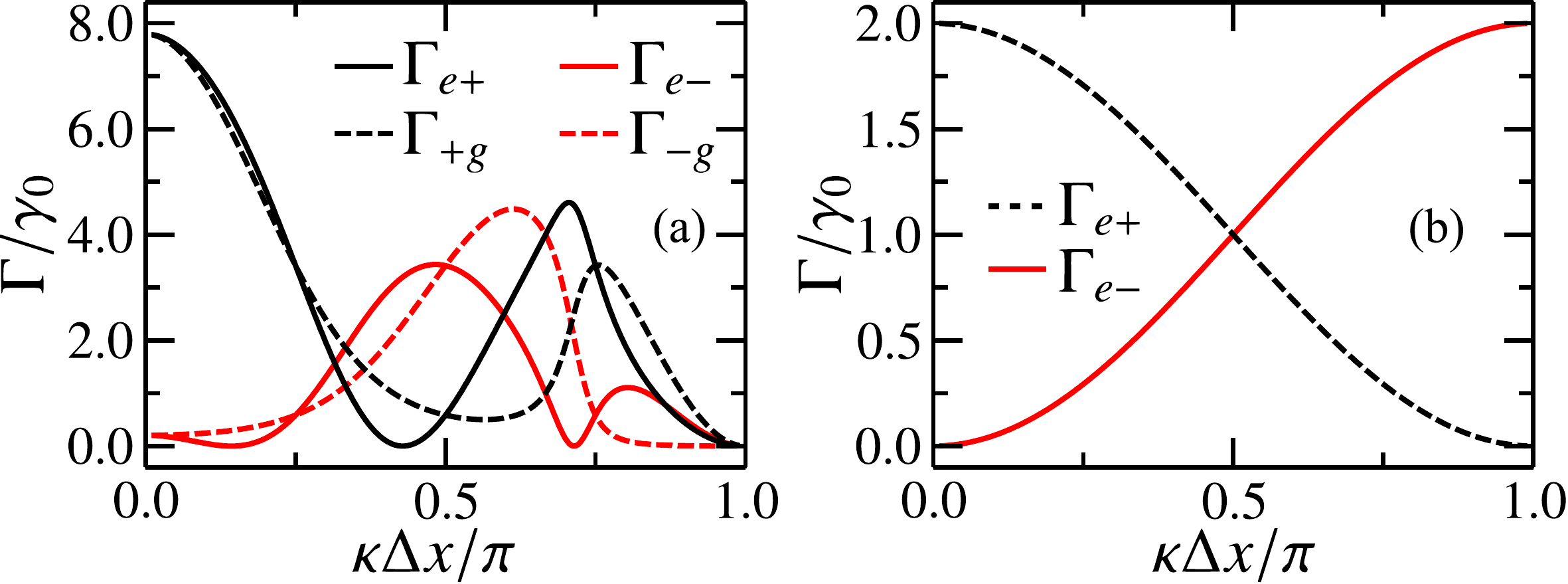}
	\caption{Collective decay rates, as a function of $\kappa \Delta x$, for (a) two nested giant atoms and (b) two small atoms. For small atoms, one has $\Gamma_{e+}\!=\!\Gamma_{+g}$ and $\Gamma_{e-}\!=\!\Gamma_{-g}$ for any $\kappa \Delta x$.}
	\label{Fig2} 
\end{figure}

Let us now focus on the particular case of two giant atoms with the same resonant frequency $\omega_{n}=\omega$ and same local relaxation rates $\gamma^{(n)}_{m}=\gamma_{0}$, connected to the waveguide at two points ($K_{1}=K_{2}=2$). We assume that the spacing between all neighbouring connection points is the same, hereafter called $\Delta x$. Fig.~\ref{Fig1}{\color{blue}a} depicts the ``nested'' configuration, where the two connections of one atom fall in-between the connections of the other. The interaction between the atoms, mediated by the waveguide, results in a shift of the energy levels, which is illustrated in Fig.~\ref{Fig1}{\color{blue}b}. The energy of the different levels is $E_{g} =0$, $E_{e}=\hbar \left( \tilde{\omega}_{1} + \tilde{\omega}_{2} \right)$,  $E_{\pm}=\hbar ( \tilde{\omega}_{1} + \tilde{\omega}_{2} \pm \tilde{\Delta}  )/2$, with $\tilde{\Delta}\!=\!\sqrt{4\Delta^2_{12} + \left(\tilde{\omega}_{1} - \tilde{\omega}_{2}\right)^2}$, which are associated to the states $\ket{gg}$, $\ket{ee}$ and
\begin{align}
	\ket{\psi_{\pm}} = \frac{1}{\Ncal_{\pm}} \left[ \frac{\tilde{\omega}_{1} - \tilde{\omega}_{2}\pm\tilde{\Delta}}{\Delta_{12}} \ket{eg} + 2 \ket{ge} \right],
\end{align}
respectively. We have introduced the normalization factor $\Ncal_{\pm}^2\!=\!4 + (\tilde{\omega}_{1} - \tilde{\omega}_{2} \pm \tilde{\Delta})^2/\Delta_{12}^2$ and the frequencies $\tilde{\omega}_{n}\!=\!\omega+\delta_{n}$. The eigenstates $\ket{\psi_{\pm}}$ are single-excitation (entangled) states, with an energy difference $\tilde{\Delta}$. The transition rates between these levels are given by~\cite{SupInf_GiantAtoms}:
\begin{subequations}
	\label{Eq-DecayRates}
	\begin{align}
		\Gamma_{e+} &= \frac{\Gamma_{2} \eta_{+} - \Gamma_{1} \eta_{-} + \xi}{2\tilde{\Delta}}, ~~
		\Gamma_{+g} = \frac{\Gamma_{1} \eta_{+} - \Gamma_{2} \eta_{-} + \xi}{2\tilde{\Delta}} , \\
		\Gamma_{e-} &= \frac{\Gamma_{1} \eta_{+} - \Gamma_{2} \eta_{-} - \xi }{2\tilde{\Delta}} , ~~
		\Gamma_{-g} = \frac{\Gamma_{2} \eta_{+} - \Gamma_{1} \eta_{-} - \xi }{2\tilde{\Delta}} ,
	\end{align}
\end{subequations}
with $\eta_{\pm}\!=\!\tilde{\omega}_{1} - \tilde{\omega}_{2} \pm \tilde{\Delta}$ and $\xi\!=\!4\Delta_{12} \Gamma_{12}$. In Fig.~\ref{Fig2} we show the behavior of each decay rate for the nested giant atoms and small ones as function of $\Delta x$.

For point-like atoms, the shift $\delta_{n}$ vanishes and all single-atom decay rates are the same: $\Gamma_{1}=\Gamma_{2}$. The shift is then $\tilde{\Delta}=2|\Gamma_{12}|$, and one obtains that each decay branch (that is, passing either through $\ket{\psi_+}$ or $\ket{\psi_-}$) possesses a single decay rate: $\Gamma_{e\pm}=\Gamma_{\pm g}=\Gamma_{2} \pm \Gamma_{12}\Delta_{12}/|\Delta_{12}|$. Also, in this case $\ket{\psi_\pm}$ are simply the symmetric and antisymmetric states~\cite{Dicke:54,Tanas-Ficek:04}. Differently, for giant atoms, even when the decay rates of all connection points are equal, $\gamma_n^{(k)}=\gamma_0$, the single atom decay rates are different for each atom ($\Gamma_1\neq\Gamma_2$) since the relative phases between their connection points are different. Furthermore, the four-level energy structure is characterized by four distinct decay rates, whereas in the case of point-like particles, there are only two distinct decay rates. As we shall now see, this complex internal structure of the two-giant-atom system leads to new regimes which cannot be reached for small atoms.

\emph{Maximally entangled steady state --} Generating stationary entanglement with small atoms coupled to common radiation modes can be achieved using energy shifts, either mediated by the interactions~\cite{Cidrim:20,Williamson:20,Trebbia:22} or between the raw transition frequencies of the emitters~\cite{Oliveira:22}. 
For giant atoms, the energy shifts are modest ($\tilde\Delta\leq 4\gamma_0$), yet 
the distinct decay rates which connect the $\ket{\psi_\pm}$ states to the fundamental and fully-excited state allows one to generate highly-entangled stationary states.

Let us illustrate this point by considering two atoms initially in the ground state, before the pump is switched on at time $t=0$. The dynamics of the populations of states $\ket{gg}$ and $\ket{\beta}=(\ket{ge}-\ket{eg})/\sqrt{2}$ (which is the maximally entangled anti-symmetric state for small atoms~\footnote{Note that this state can be written as a linear combination of the sub- and sub-radiant modes $\ket{\psi_{\pm}}$ introduced before for giant atoms.}) for small and giant atoms separated by a distance $\kappa\Delta x=0.01\pi$ is shown in Fig.~\ref{Fig1}{\color{blue}c}. Remarkably, for giant atoms, although the energy shifts are negligible ($\tilde\Delta\approx2\sqrt{10}\gamma_{0}\kappa\Delta x\ll\gamma_0$), the system reaches a steady state where almost all the population in the state $\ket{\beta}$, with a long lifetime $1/\Gamma_-\approx 546/\gamma_{0}$~\cite{SupInf_GiantAtoms}. This highly entangled steady-state is represented as a density matrix in Fig.~\ref{Fig1}{\color{blue}d}, where the population is concentrated on the single-excitation sector. Oppositely, for small atoms the system is driven toward a superposition of the different states, including the ground state, and it does not become entangled.

We here quantify the entanglement using the concurrence of the state, as proposed by Hill and Wootters for a pair of two-level system~\cite{Hill:97}. For example, the maximally-entangled state $\ket{\beta}$ introduced above reaches the value $\Ccal(\rho)=1$, whereas it is zero for non-entangled states~\cite{Nielsen:Book}. In the particular case discussed above, see Fig.~\ref{Fig1}{\color{blue}c-d}, giant atoms present a concurrence of $\Ccal=0.995$ in the steady-state, while it is precisely $\Ccal=0$ for small atoms.

\begin{figure}[t!]
	\centering
	\includegraphics[width=\columnwidth]{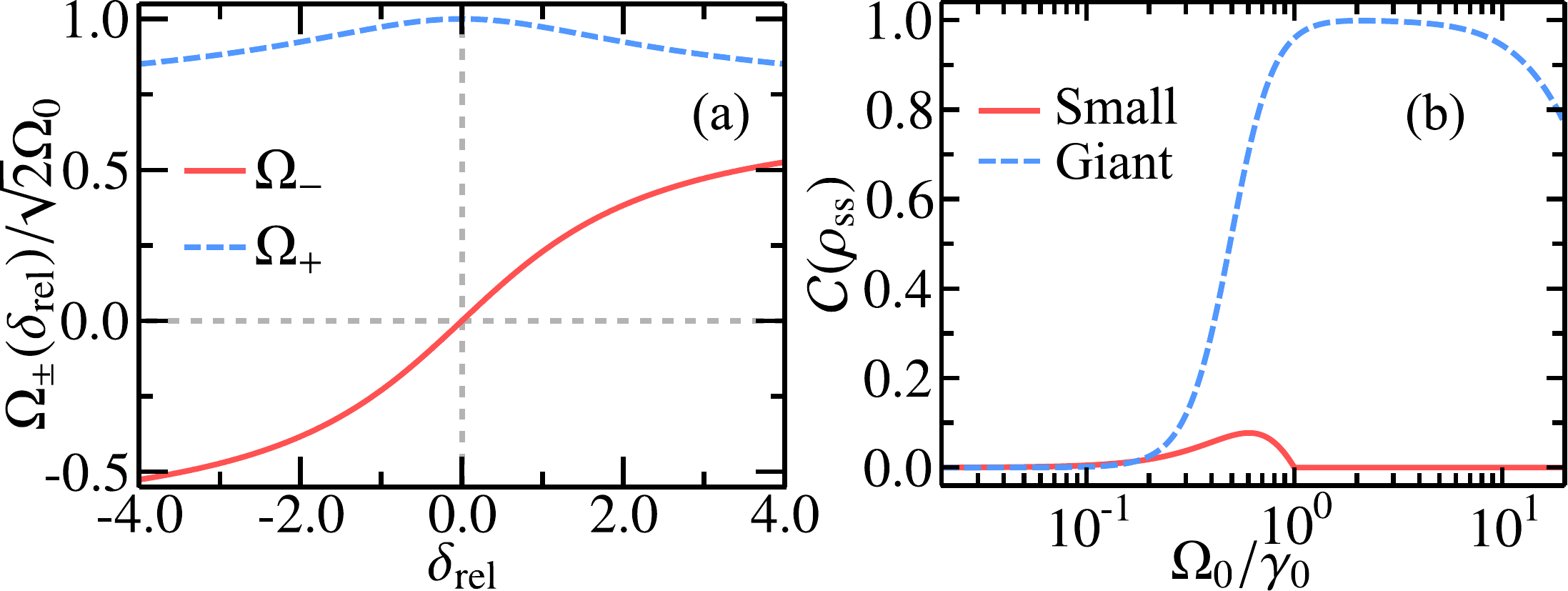}
	\caption{(a) Effective coupling as function of the relative energy shift $\delta_{\text{rel}}$ for two nested giant atoms, written as a multiple of the maximum effective coupling for two identical small atoms $\sqrt{2}\Omega_{0}$. (b) Steady-state entanglement obtained through local (resonant) fields driving two nested giant atoms, where the wavenumber $\kappa$ is considered so that $\kappa \Delta x\!=\!0.01\pi$.}
	\label{Fig3} 
\end{figure}

Let us now discuss the mechanism behind the efficient generation of steady-state entanglement. 
The efficient coupling of the external field with the atomic transition $\ket{gg}\!\rightarrow\!\ket{\psi_{-}}$ plays an important role in this process. The coupling induced by the driving field between the ground state and $\ket{\psi_{\pm}}$ can be obtained by rewriting the pump term $H_{\text{ext}}=\hbar(\Omega_+\hat{\sigma}^+_+ +\Omega_-\hat{\sigma}^+_-)+h.c.$, with $\hat{\sigma}^+_\pm=\ket{\psi_{\pm}}\bra{gg}$, and the effective coupling coefficients $\Omega_{\pm}$ given by
\begin{align}
	\Omega_{\pm} = \Omega_{0} \left(\frac{\delta_{12} + 2\Delta_{12} \pm \sqrt{4\Delta_{12}^2 + \delta_{12}^2 } }{\sqrt{8\Delta_{12}^2 + 2\delta_{12} \left( \delta_{12} \pm \sqrt{4\Delta_{12}^2 + \delta_{12}^2 } \right)  }}\right) ,
\end{align}
with $\delta_{12}\!=\!\delta_{1}-\delta_{2}$. As for small atoms, the driving field allows for population transfer from $\ket{gg}$ to $\ket{ee}$ through the state $\ket{\psi_{+}}$ at a rate $\Omega_{+}$. However, we highlight the effective coupling value of $\Omega_{-}$, which leads to a direct coupling of the transition from the state $\ket{gg}$ to $\ket{\psi_{-}}$ in a coherent way. By defining the relative phase shift $\delta_{\text{rel}}\!=\!\delta_{12}/\Delta_{12}$, we observe that for $\delta_{\text{rel}}\!\neq\!0$, one has $\Omega_{-}\!\neq\!0$ (see Fig.~\ref{Fig3}{\color{blue}a}). In particular, whenever $\delta_{1}\!=\!\delta_{2}$ we get $\Omega_{-}\!=\!0$ and $\Omega_{+}\!=\!\sqrt{2} \Omega_{0}$, as observed for small atoms. In this sense, the difference in resonant energy of the atoms due to their finite size induces an efficient coupling to the $\ket{\psi_{-}}$ state.

Note that this mechanism alone is not enough to get a large population in the state $\ket{\psi_{-}}$, since a non-zero coupling between $\ket{gg}$ and $\ket{\psi_{-}}$ also can be achieved for small atoms with different frequencies $\omega_{1}$ and $\omega_{2}$~\cite{SupInf_GiantAtoms}. The interference between the contributions of the states $\ket{\psi_{\pm}}$ in the steady state needs also to be taken into account, since the maximally entangled steady-state $\ket{\beta}$ can is written as superposition of these two states. The generation of steady-state entanglement is optimized when the pump is strong enough for a single-excitation state to be fully populated, yet not enough to populate all the state in the system. As illustrated in Fig.~\ref{Fig3}{\color{blue}b} the concurrence reaches its maximum when external pump has a Rabi frequency $\Omega_{0}\approx 2\gamma_0$. While maximally entangled states are not possible with same-energy small atoms, with a maximum value $\max_{\Omega_{0}}[\Ccal(\rho_{\text{ss}}^{\text{small}})]\!<\!0.1$, giant atoms allow one to achieve highly entangled states, with concurrence of order of $\Ccal(\rho_{\text{ss}}^{\text{nested}})\!\approx\!0.999$. As one can prepare any of the other maximally entangled Bell state from $\ket{\beta}$ by single atom (local) operations~\cite{Nielsen:Book}, the steady-state approach proposed in this work can be useful to create arbitrary entangled state of the two giant atoms.

\begin{figure}[t!]
	\centering
	\includegraphics[width=\columnwidth]{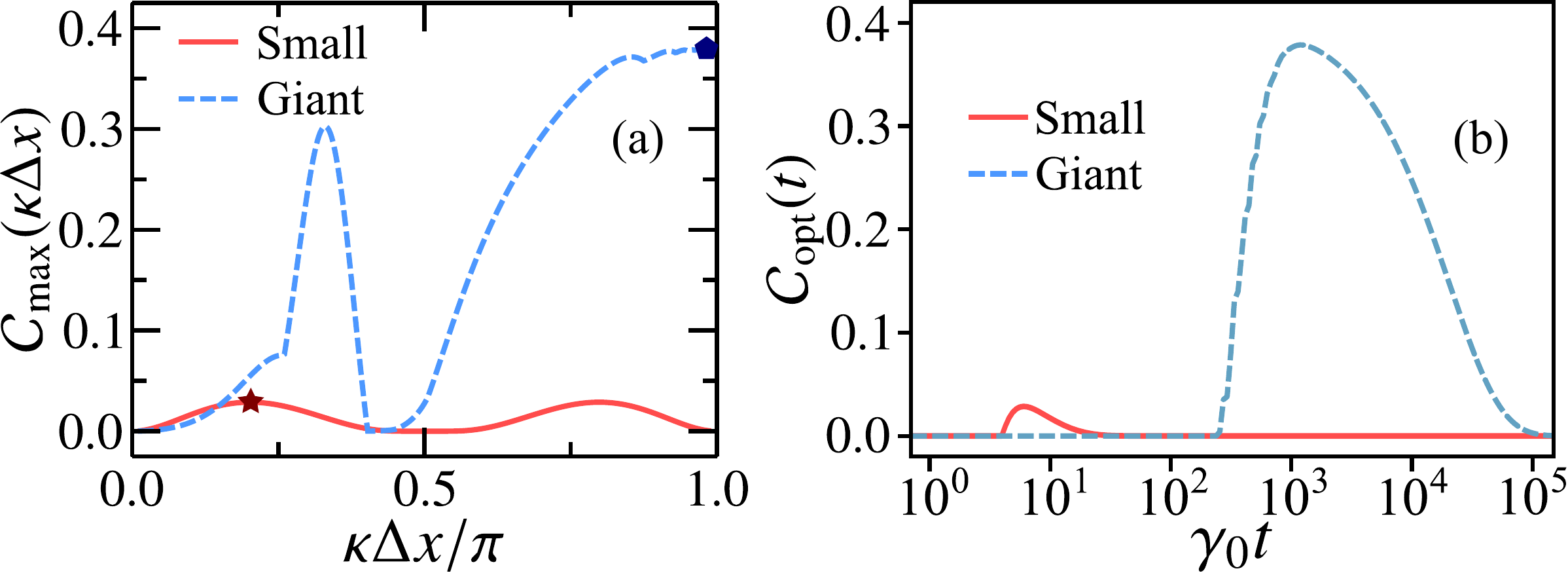}
	\caption{(a) Maximum amount of entanglement generated through the decay process by sudden-birth for small and two giant nested atoms as a function of $\kappa\Delta x$. (b) Dynamics of the optimal sudden-birth of entanglement for small and giant nested atoms as a function of $\gamma_{0} t$. We set $\kappa \Delta x \approx 0.19\pi$ for small atoms and $\kappa \Delta x\!=\!0.99\pi$ for the system of two giant nested atoms, the values highlighted in (a) with a star symbol.}
	\label{Decay:Fig} 
\end{figure}

\emph{Sudden birth of entanglement --} Let us now consider an initially fully-inverted state $\ket{ee}$, which can be prepared by applying a fast $\pi$-pulse on the atoms. We then let the system decay, in absence of pump field, monitoring how the action of collective spontaneous emission through sub- and super-radiant branches $\ket{\psi_{\pm}}$ allows one to efficiently generate entanglement. The system first decays toward a mixture of the $\ket{\psi_+}$ and $\ket{\psi_-}$ states, which is in general not entangled, although each of these states is individually entangled. However, because of the difference in their lifetimes $1/\Gamma_{\pm g}$, the $\ket{\psi_+}$ component quickly decays to the ground state while the $\ket{\psi_-}$ remains for a time $\sim1/\Gamma_{-g}$.  The concurrence $\Ccal_{\text{max}}$ as function of the spacing $\kappa \Delta x$ and  time, for giant and small atoms, is shown in Fig.~\ref{Decay:Fig}{\color{blue}a}, with a the maximum amount of entanglement $\Ccal_{\text{max}}(\kappa \Delta x)\!=\!\max_{t}\Ccal(\kappa \Delta x,t)$ created by decay for $\kappa \Delta x\!=\!0.99\pi$. The decay dynamics in Fig.~\ref{Decay:Fig}{\color{blue}b} reveals that this concurrence is created at late times. The difference in the entanglement generation comes from the fact that for small atoms, this entanglement comes from the population of the superradiant mode, whereas for giant atoms the entanglement results from the population in the subradiant mode~\cite{SupInf_GiantAtoms}. 

The sudden birth of entanglement in giant atoms comes from the unbalanced amount of population in the states $\sket{\psi_{\pm}}$, which arises from the asymmetry between the decay rates $\Gamma_{e\pm}$ and $\Gamma_{\pm g}$. Since $\Gamma_{e -}\gg\Gamma_{- g}$ and $\Gamma_{e +}<\Gamma_{+ g}$ (for the case highlighted in Fig.~\ref{Decay:Fig}{\color{blue}a}), the atomic population is transferred from $\sket{ee}$ to $\sket{\psi_{-}}$ faster than from $\sket{\psi_{-}}$ to $\sket{gg}$, while population decaying from $\sket{ee}$ to $\sket{\psi_{+}}$ is quickly transferred to $\sket{gg}$. It leads to an efficient generation of entanglement through the control of the waveguide wavenumber $\kappa$. In Ref.~\cite{SupInf_GiantAtoms}, we show that other geometries of giant atoms lead to a sudden birth of entanglement which is comparable to the one for small atoms (Figs.~\ref{Decay:Fig}{\color{blue}a} and~\ref{Decay:Fig}{\color{blue}b}).

\emph{Giant photon bunching in giant atoms.} Let us now discuss how the statistics of the light emitted by two giant atoms in the waveguide can be used as an entanglement witness of the system. The light emitted by the atoms is described by the electric field operator $\hat{E}_{\text{a}}(t,\vec{r})$ emitted by the atoms. For our system, we can show that the left and right travelling fields read~\cite{SupInf_GiantAtoms}
\begin{align}
	\hat{E}^{\mathbf{l}}_{\text{a}}(t) \propto \sum_{n=1}^{N}\sum_{j=1}^{K_{n}} e^{-ik x_{j}^{(n)}}\sigma_{n}^{-}(t) , ~ \hat{E}_{\text{a}}^{\mathbf{r}}(t) \propto \sum_{n=1}^{N}\sum_{j=1}^{K_{n}} e^{ik x_{j}^{(n)}}\sigma_{n}^{-}(t),
\end{align}
where we have assumed that the phase at a given connection point $x_{\ell}^{(n)}$ of the atom varies very slowly with the wavenumber $\kappa$. 

We here characterize the emitted light by its second-order correlation function $g^{(2)}(t,t +\tau)$ which, for the field $\hat{E}_{\text{a}}^{\alpha}(t)$ ($\alpha\!=\!\{\mathbf{l},\mathbf{r}\}$ for left- and right-emitted field), is given by
\begin{align}
	g^{(2)}_{\alpha}(t,t+\tau) = \frac{\langle \hat{E}_{\text{a}}^{\alpha\dagger}(t)\hat{E}_{\text{a}}^{\alpha\dagger}(t+\tau)\hat{E}_{\text{a}}^{\alpha}(t+\tau)\hat{E}_{\text{a}}^{\alpha}(t)\rangle}{\langle \hat{E}_{\text{a}}^{\alpha\dagger}(t)\hat{E}_{\text{a}}^{\alpha}(t)\rangle\langle\hat{E}_{\text{a}}^{\alpha\dagger}(t+\tau)\hat{E}_{\text{a}}^{\alpha}(t+\tau)\rangle}.
\end{align}
The steady-state limit $g^{(2)}_{\alpha}(\tau)\!=\!\lim_{t\rightarrow\infty} g^{(2)}_{\alpha}(t,t+\tau)$ is considered. Finally, we define the Mandel $Q$-parameter for the field $\hat{E}^{\alpha}_{\text{a}}(t,\vec{r})$ given by~\cite{Mandel:79}
\begin{align}
	Q_{\alpha} =\lim_{t\rightarrow\infty} \langle \hat{E}^{\alpha\dagger}_{\text{a}}(t) \hat{E}_{\text{a}}^{\alpha}(t)\rangle \left( g^{(2)}_{\alpha}(t,t) - 1 \right) ,
\end{align}
where the limit in $t$ refers to the steady-state limit. 

\begin{figure}[t!]
	\centering
	\includegraphics[width=\columnwidth]{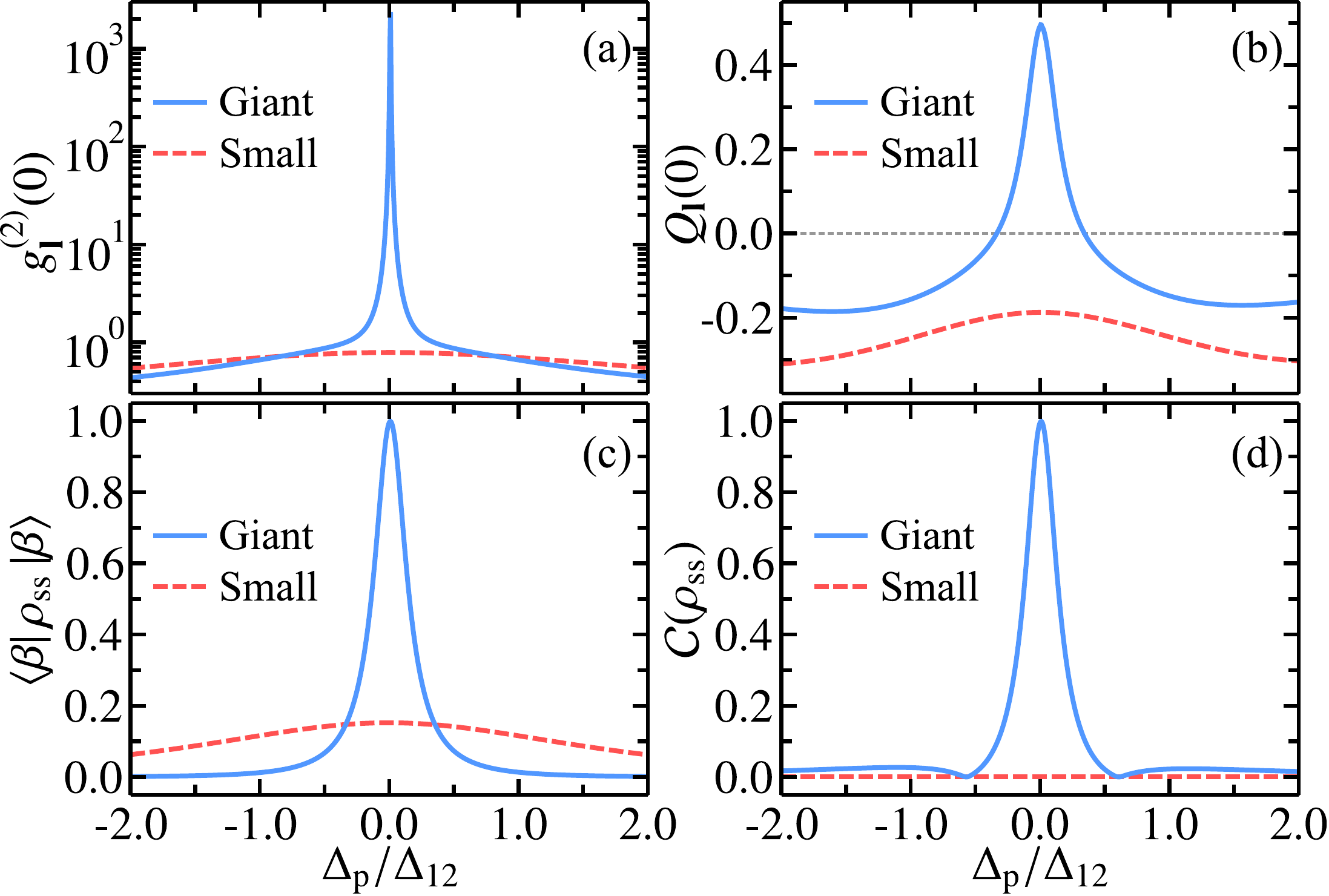}
	\caption{Behavior of (a) the function $g^{(2)}_{\alpha}(0)$ and (b) the Mandel parameter, for the emitted light by small and giant atoms, as function of the pumping detuning $\Delta_{\text{p}}$. (c) The population in the maximally entangled state $\ket{\beta}$ and (d) the corresponding concurrence. Here we used $\Omega_{0}\!=\!1.5\gamma_{0}$ and $\kappa \Delta x\!=\!0.01\pi$.}
	\label{Fig4} 
\end{figure}

When we tune the frequency of the external field with respect to the atomic transition $\omega$, $\omega_{\text{field}}\!=\!\omega-\Delta_{\text{p}}$, the atomic steady state, and consequently the statistics of the emitted light, change as shown in Fig.~\ref{Fig4}. The functions $g^{(2)}_{\alpha}(0)$ and $Q_{\alpha}$ are shown in Figs.~\ref{Fig4}{\color{blue}a} and~\ref{Fig4}{\color{blue}b}, respectively. The light statistics can be changed from sub-Poissonian statistics to super-bunching by adjusting the detuning $\Delta_{\text{p}}$ in the interval $[-2\Delta_{12},2\Delta_{12}]$. We stress that while the light statistics of small atoms remain sub-Poissonian, the light emitted from giant atoms exhibit different property depending on the detuning $\Delta_{\text{p}}$. In particular, giant photon bunching has been reported in a system of quantum dots~\cite{Jahnke:16}, a system which in principle may behave as giant atoms due to the spatial extent of the system, as compared to the pump wavelength~\cite{Arcari:14,Kockum:20,Gines:22}.

In Figs.~\ref{Fig4}{\color{blue}c} and~\ref{Fig4}{\color{blue}d} we show the entanglement and the population of the long-lived state $\ket{\beta}$, respectively. From Fig.~\ref{Fig4}{\color{blue}c}, one can observe that when the external pump is at resonance, $\Delta_{\text{p}}\!\approx\!0$, the steady state of the system is given by $\rho_{\text{ss}}^{\text{nested}}\!\approx\!\ket{\beta}\bra{\beta}$, such that the entanglement originates due to the large population of the long-lived maximally entangled state $\ket{\beta}$. In this scenario, the similar behavior of the coherence second-order function and Mandel parameter suggests that these quantities work as witness of maximally entangled states of the system. In other words, although bunching of light is not a measurement/witness of entanglement in general, the giant bunching observed in the steady-state regime of our dynamics is here associated with the generation of long-lived maximally entangled states for two nested giant atoms. In this particular setup, the photon statistics could thus be used to monitor the atomic state.

\emph{Conclusion.} Considering a system of two giant atoms driven by photon-mediated interaction and external fields, we have studied entanglement generation, and quantified it by the concurrence between the atoms. We showed that the phase acquired by the traveling photon inside the waveguide (related to the frequency of the light in the waveguide~\cite{Kockum:14}) allows us to design a two-qubit system with both frequency-tunable collective decay rates and transition rates between the states of the system. Consequently, frequency-tunable collective effects of two giant atoms can be achieved. We have taken advantage of these properties to create a high degree of entanglement in the system through two different processes: free decay and pump-driven dynamics. The amount of concurrence obtained for giant atoms is much larger than for small (pointlike) atoms. In particular, it is possible to create a quasi-maximally-entangled steady state with giant atoms, by adequately adjusting the Rabi frequency of the classical external field which drives the system. Finally, we observe that the statistics of the light emitted in the waveguide are a witness of the concurrence generated in the system, so it can be used as a non-destructive measurement for entanglement. By adjusting the external field detuning $\Delta_{\text{p}}$, giant photon bunching emission~\cite{Jahnke:16} is observed as the entanglement increases.

Our results can be implemented in superconducting quantum circuits, since the experimental realization of a similar system has been reported by Kannan \textit{et al}~\cite{Kannan:20}. To this end, the two frequency-tunable transmon qubits used in Ref.~\cite{Kannan:20} should be coupled to the transmission line coplanar waveguide at two locations as shown in Fig.~\ref{Fig1}{\color{blue}a}, instead the braided manner considered in Ref.~\cite{Kannan:20}. Therefore, experimental realization and the immediate extension of our analysis for systems of many giant atoms are left for future prospects.

\begin{acknowledgements}
	A. C. S. and R. B. are supported by São Paulo Research Foundation (FAPESP) (Grant No 2018/15554-5, 2019/13143-0, 2019/22685-1 and 2021/10224-0). R. B. has received support from grants from the National Council for Scientific and Technological Development (CNPq, Grant Nos. 313886/2020-2 and 409946/2018-4). This work has been supported by the French government, through the UCA$^\textrm{JEDI}$ Investments in the Future project managed by the National Research Agency (ANR) with the reference number ANR-15-IDEX-01.
\end{acknowledgements}


\begin{thebibliography}{51}%
	\makeatletter
	\providecommand \@ifxundefined [1]{%
		\@ifx{#1\undefined}
	}%
	\providecommand \@ifnum [1]{%
		\ifnum #1\expandafter \@firstoftwo
		\else \expandafter \@secondoftwo
		\fi
	}%
	\providecommand \@ifx [1]{%
		\ifx #1\expandafter \@firstoftwo
		\else \expandafter \@secondoftwo
		\fi
	}%
	\providecommand \natexlab [1]{#1}%
	\providecommand \enquote  [1]{``#1''}%
	\providecommand \bibnamefont  [1]{#1}%
	\providecommand \bibfnamefont [1]{#1}%
	\providecommand \citenamefont [1]{#1}%
	\providecommand \href@noop [0]{\@secondoftwo}%
	\providecommand \href [0]{\begingroup \@sanitize@url \@href}%
	\providecommand \@href[1]{\@@startlink{#1}\@@href}%
	\providecommand \@@href[1]{\endgroup#1\@@endlink}%
	\providecommand \@sanitize@url [0]{\catcode `\\12\catcode `\$12\catcode
		`\&12\catcode `\#12\catcode `\^12\catcode `\_12\catcode `\%12\relax}%
	\providecommand \@@startlink[1]{}%
	\providecommand \@@endlink[0]{}%
	\providecommand \url  [0]{\begingroup\@sanitize@url \@url }%
	\providecommand \@url [1]{\endgroup\@href {#1}{\urlprefix }}%
	\providecommand \urlprefix  [0]{URL }%
	\providecommand \Eprint [0]{\href }%
	\providecommand \doibase [0]{http://dx.doi.org/}%
	\providecommand \selectlanguage [0]{\@gobble}%
	\providecommand \bibinfo  [0]{\@secondoftwo}%
	\providecommand \bibfield  [0]{\@secondoftwo}%
	\providecommand \translation [1]{[#1]}%
	\providecommand \BibitemOpen [0]{}%
	\providecommand \bibitemStop [0]{}%
	\providecommand \bibitemNoStop [0]{.\EOS\space}%
	\providecommand \EOS [0]{\spacefactor3000\relax}%
	\providecommand \BibitemShut  [1]{\csname bibitem#1\endcsname}%
	\let\auto@bib@innerbib\@empty
	\bibitem [{\citenamefont {Wallraff}\ \emph {et~al.}(2004)\citenamefont
		{Wallraff}, \citenamefont {Schuster}, \citenamefont {Blais}, \citenamefont
		{Frunzio}, \citenamefont {Huang}, \citenamefont {Majer}, \citenamefont
		{Kumar}, \citenamefont {Girvin},\ and\ \citenamefont
		{Schoelkopf}}]{wallraff2004}%
	\BibitemOpen
	\bibfield  {author} {\bibinfo {author} {\bibfnamefont {A.}~\bibnamefont
			{Wallraff}},  {\em et~al.},\ }\enquote{\bibinfo {title} {Strong coupling of a
			single photon to a superconducting qubit using circuit quantum
			electrodynamics}},\ \href {https://www.nature.com/articles/nature02851}
	{\bibfield  {journal} {\bibinfo  {journal} {Nature}\ }\textbf {\bibinfo
			{volume} {431}},\ \bibinfo {pages} {162} (\bibinfo {year}
		{2004})}\BibitemShut {NoStop}%
	\bibitem [{\citenamefont {van Loo}\ \emph {et~al.}(2013)\citenamefont {van
			Loo}, \citenamefont {Fedorov}, \citenamefont {Lalumière}, \citenamefont
		{Sanders}, \citenamefont {Blais},\ and\ \citenamefont {Wallraff}}]{Arjan:13}%
	\BibitemOpen
	\bibfield  {author} {\bibinfo {author} {\bibfnamefont {A.~F.}\ \bibnamefont
			{van Loo}},  {\em et~al.},\ }\enquote{\bibinfo {title} {Photon-Mediated
			Interactions Between Distant Artificial Atoms}},\ \href {\doibase
		10.1126/science.1244324} {\bibfield  {journal} {\bibinfo  {journal}
			{Science}\ }\textbf {\bibinfo {volume} {342}},\ \bibinfo {pages} {1494}
		(\bibinfo {year} {2013})}\BibitemShut {NoStop}%
	\bibitem [{\citenamefont {Mlynek}\ \emph {et~al.}(2014)\citenamefont {Mlynek},
		\citenamefont {Abdumalikov}, \citenamefont {Eichler},\ and\ \citenamefont
		{Wallraff}}]{Mlynek:14}%
	\BibitemOpen
	\bibfield  {author} {\bibinfo {author} {\bibfnamefont {J.~A.}\ \bibnamefont
			{Mlynek}}, \bibinfo {author} {\bibfnamefont {A.~A.}\ \bibnamefont
			{Abdumalikov}}, \bibinfo {author} {\bibfnamefont {C.}~\bibnamefont
			{Eichler}},  and \bibinfo {author} {\bibfnamefont {A.}~\bibnamefont
			{Wallraff}},\ }\enquote{\bibinfo {title} {Observation of Dicke superradiance
			for two artificial atoms in a cavity with high decay rate}},\ \href {\doibase
		https://doi.org/10.1038/ncomms6186} {\bibfield  {journal} {\bibinfo
			{journal} {Nature communications}\ }\textbf {\bibinfo {volume} {5}},\
		\bibinfo {pages} {1} (\bibinfo {year} {2014})}\BibitemShut {NoStop}%
	\bibitem [{\citenamefont {Shen}\ and\ \citenamefont {Fan}(2007)}]{Shen:07}%
	\BibitemOpen
	\bibfield  {author} {\bibinfo {author} {\bibfnamefont {J.-T.}\ \bibnamefont
			{Shen}} and \bibinfo {author} {\bibfnamefont {S.}~\bibnamefont {Fan}},\
	}\enquote{\bibinfo {title} {Strongly Correlated Two-Photon Transport in a
			One-Dimensional Waveguide Coupled to a Two-Level System}},\ \href {\doibase
		10.1103/PhysRevLett.98.153003} {\bibfield  {journal} {\bibinfo  {journal}
			{Phys. Rev. Lett.}\ }\textbf {\bibinfo {volume} {98}},\ \bibinfo {pages}
		{153003} (\bibinfo {year} {2007})}\BibitemShut {NoStop}%
	\bibitem [{\citenamefont {Zheng}\ \emph {et~al.}(2012)\citenamefont {Zheng},
		\citenamefont {Gauthier},\ and\ \citenamefont {Baranger}}]{Zheng:12}%
	\BibitemOpen
	\bibfield  {author} {\bibinfo {author} {\bibfnamefont {H.}~\bibnamefont
			{Zheng}}, \bibinfo {author} {\bibfnamefont {D.~J.}\ \bibnamefont {Gauthier}},
		and \bibinfo {author} {\bibfnamefont {H.~U.}\ \bibnamefont {Baranger}},\
	}\enquote{\bibinfo {title} {Strongly correlated photons generated by coupling
			a three- or four-level system to a waveguide}},\ \href {\doibase
		10.1103/PhysRevA.85.043832} {\bibfield  {journal} {\bibinfo  {journal} {Phys.
				Rev. A}\ }\textbf {\bibinfo {volume} {85}},\ \bibinfo {pages} {043832}
		(\bibinfo {year} {2012})}\BibitemShut {NoStop}%
	\bibitem [{\citenamefont {Wang}\ \emph {et~al.}(2020)\citenamefont {Wang},
		\citenamefont {Li}, \citenamefont {Feng}, \citenamefont {Song}, \citenamefont
		{Song}, \citenamefont {Liu}, \citenamefont {Guo}, \citenamefont {Zhang},
		\citenamefont {Dong}, \citenamefont {Zheng}, \citenamefont {Wang},\ and\
		\citenamefont {Wang}}]{Wang:20}%
	\BibitemOpen
	\bibfield  {author} {\bibinfo {author} {\bibfnamefont {Z.}~\bibnamefont
			{Wang}},  {\em et~al.},\ }\enquote{\bibinfo {title} {Controllable Switching
			between Superradiant and Subradiant States in a 10-qubit Superconducting
			Circuit}},\ \href {\doibase 10.1103/PhysRevLett.124.013601} {\bibfield
		{journal} {\bibinfo  {journal} {Phys. Rev. Lett.}\ }\textbf {\bibinfo
			{volume} {124}},\ \bibinfo {pages} {013601} (\bibinfo {year}
		{2020})}\BibitemShut {NoStop}%
	\bibitem [{\citenamefont {Gottesman}\ and\ \citenamefont
		{Chuang}(1999)}]{Gottesman:99}%
	\BibitemOpen
	\bibfield  {author} {\bibinfo {author} {\bibfnamefont {D.}~\bibnamefont
			{Gottesman}} and \bibinfo {author} {\bibfnamefont {I.~L.}\ \bibnamefont
			{Chuang}},\ }\enquote{\bibinfo {title} {Demonstrating the viability of
			universal quantum computation using teleportation and single-qubit
			operations}},\ \href {\doibase https://doi.org/10.1038/46503} {\bibfield
		{journal} {\bibinfo  {journal} {Nature}\ }\textbf {\bibinfo {volume} {402}},\
		\bibinfo {pages} {390} (\bibinfo {year} {1999})}\BibitemShut {NoStop}%
	\bibitem [{\citenamefont {DiCarlo}\ \emph {et~al.}(2009)\citenamefont
		{DiCarlo}, \citenamefont {Chow}, \citenamefont {Gambetta}, \citenamefont
		{Bishop}, \citenamefont {Johnson}, \citenamefont {Schuster}, \citenamefont
		{Majer}, \citenamefont {Blais}, \citenamefont {Frunzio}, \citenamefont
		{Girvin} \emph {et~al.}}]{dicarlo2010}%
	\BibitemOpen
	\bibfield  {author} {\bibinfo {author} {\bibfnamefont {L.}~\bibnamefont
			{DiCarlo}},  {\em et~al.},\ }\enquote{\bibinfo {title} {Demonstration of
			two-qubit algorithms with a superconducting quantum processor}},\ \href
	{https://www.nature.com/articles/nature08121} {\bibfield  {journal} {\bibinfo
			{journal} {Nature}\ }\textbf {\bibinfo {volume} {460}},\ \bibinfo {pages}
		{240} (\bibinfo {year} {2009})}\BibitemShut {NoStop}%
	\bibitem [{\citenamefont {Bennett}\ \emph {et~al.}(1993)\citenamefont
		{Bennett}, \citenamefont {Brassard}, \citenamefont {Cr\'epeau}, \citenamefont
		{Jozsa}, \citenamefont {Peres},\ and\ \citenamefont {Wootters}}]{Bennett:93}%
	\BibitemOpen
	\bibfield  {author} {\bibinfo {author} {\bibfnamefont {C.~H.}\ \bibnamefont
			{Bennett}},  {\em et~al.},\ }\enquote{\bibinfo {title} {Teleporting an
			unknown quantum state via dual classical and Einstein-Podolsky-Rosen
			channels}},\ \href {\doibase 10.1103/PhysRevLett.70.1895} {\bibfield
		{journal} {\bibinfo  {journal} {Phys. Rev. Lett.}\ }\textbf {\bibinfo
			{volume} {70}},\ \bibinfo {pages} {1895} (\bibinfo {year}
		{1993})}\BibitemShut {NoStop}%
	\bibitem [{\citenamefont {{Ren}}\ \emph {et~al.}(2017)\citenamefont {{Ren}},
		\citenamefont {{Xu}}, \citenamefont {{Yong}}, \citenamefont {{Zhang}},
		\citenamefont {{Liao}}, \citenamefont {{Yin}}, \citenamefont {{Liu}},
		\citenamefont {{Cai}}, \citenamefont {{Yang}}, \citenamefont {{Li}},
		\citenamefont {{Yang}}, \citenamefont {{Han}}, \citenamefont {{Yao}},
		\citenamefont {{Li}}, \citenamefont {{Wu}}, \citenamefont {{Wan}},
		\citenamefont {{Liu}}, \citenamefont {{Liu}}, \citenamefont {{Kuang}},
		\citenamefont {{He}}, \citenamefont {{Shang}}, \citenamefont {{Guo}},
		\citenamefont {{Zheng}}, \citenamefont {{Tian}}, \citenamefont {{Zhu}},
		\citenamefont {{Liu}}, \citenamefont {{Lu}}, \citenamefont {{Shu}},
		\citenamefont {{Chen}}, \citenamefont {{Peng}}, \citenamefont {{Wang}},\ and\
		\citenamefont {{Pan}}}]{Ren:17}%
	\BibitemOpen
	\bibfield  {author} {\bibinfo {author} {\bibfnamefont {J.-G.}\ \bibnamefont
			{{Ren}}},  {\em et~al.},\ }\enquote{\bibinfo {title} {Ground-to-satellite
			quantum teleportation}},\ \href {\doibase 10.1038/nature23675} {\bibfield
		{journal} {\bibinfo  {journal} {Nature}\ }\textbf {\bibinfo {volume} {549}},\
		\bibinfo {pages} {70} (\bibinfo {year} {2017})}\BibitemShut {NoStop}%
	\bibitem [{\citenamefont {Tacchino}\ \emph {et~al.}(2018)\citenamefont
		{Tacchino}, \citenamefont {Auff\`eves}, \citenamefont {Santos},\ and\
		\citenamefont {Gerace}}]{Tacchino:18}%
	\BibitemOpen
	\bibfield  {author} {\bibinfo {author} {\bibfnamefont {F.}~\bibnamefont
			{Tacchino}}, \bibinfo {author} {\bibfnamefont {A.}~\bibnamefont
			{Auff\`eves}}, \bibinfo {author} {\bibfnamefont {M.~F.}\ \bibnamefont
			{Santos}},  and \bibinfo {author} {\bibfnamefont {D.}~\bibnamefont
			{Gerace}},\ }\enquote{\bibinfo {title} {Steady State Entanglement beyond
			Thermal Limits}},\ \href {\doibase 10.1103/PhysRevLett.120.063604} {\bibfield
		{journal} {\bibinfo  {journal} {Phys. Rev. Lett.}\ }\textbf {\bibinfo
			{volume} {120}},\ \bibinfo {pages} {063604} (\bibinfo {year}
		{2018})}\BibitemShut {NoStop}%
	\bibitem [{\citenamefont {Wang}\ \emph {et~al.}(2019)\citenamefont {Wang},
		\citenamefont {Wu},\ and\ \citenamefont {Wang}}]{Wang:PRA19}%
	\BibitemOpen
	\bibfield  {author} {\bibinfo {author} {\bibfnamefont {Z.}~\bibnamefont
			{Wang}}, \bibinfo {author} {\bibfnamefont {W.}~\bibnamefont {Wu}},  and
		\bibinfo {author} {\bibfnamefont {J.}~\bibnamefont {Wang}},\
	}\enquote{\bibinfo {title} {Steady-state entanglement and coherence of two
			coupled qubits in equilibrium and nonequilibrium environments}},\ \href
	{\doibase 10.1103/PhysRevA.99.042320} {\bibfield  {journal} {\bibinfo
			{journal} {Phys. Rev. A}\ }\textbf {\bibinfo {volume} {99}},\ \bibinfo
		{pages} {042320} (\bibinfo {year} {2019})}\BibitemShut {NoStop}%
	\bibitem [{\citenamefont {Huelga}\ \emph {et~al.}(2012)\citenamefont {Huelga},
		\citenamefont {Rivas},\ and\ \citenamefont {Plenio}}]{Huelga:12}%
	\BibitemOpen
	\bibfield  {author} {\bibinfo {author} {\bibfnamefont {S.~F.}\ \bibnamefont
			{Huelga}}, \bibinfo {author} {\bibfnamefont {A.}~\bibnamefont {Rivas}},  and
		\bibinfo {author} {\bibfnamefont {M.~B.}\ \bibnamefont {Plenio}},\
	}\enquote{\bibinfo {title} {Non-Markovianity-Assisted Steady State
			Entanglement}},\ \href {\doibase 10.1103/PhysRevLett.108.160402} {\bibfield
		{journal} {\bibinfo  {journal} {Phys. Rev. Lett.}\ }\textbf {\bibinfo
			{volume} {108}},\ \bibinfo {pages} {160402} (\bibinfo {year}
		{2012})}\BibitemShut {NoStop}%
	\bibitem [{\citenamefont {{Oliveira}}\ \emph {et~al.}(2022)\citenamefont
		{{Oliveira}}, \citenamefont {{Higgins}}, \citenamefont {{Zhang}},
		\citenamefont {{Predojevi{\'c}}}, \citenamefont {{Hennrich}}, \citenamefont
		{{Bachelard}},\ and\ \citenamefont {{Villas-Boas}}}]{Oliveira:22}%
	\BibitemOpen
	\bibfield  {author} {\bibinfo {author} {\bibfnamefont {M.~H.}\ \bibnamefont
			{{Oliveira}}},  {\em et~al.},\ }\enquote{\bibinfo {title} {{Steady-state
				entanglement generation for non-degenerate qubits}}},\ \href@noop {}
	{\bibfield  {journal} {\bibinfo  {journal} {arXiv e-prints}\ ,\ \bibinfo
			{eid} {arXiv:2205.10590}} (\bibinfo {year} {2022})},\ \Eprint
	{http://arxiv.org/abs/2205.10590} {arXiv:2205.10590 [quant-ph]} \BibitemShut
	{NoStop}%
	\bibitem [{\citenamefont {Frisk~Kockum}\ \emph {et~al.}(2014)\citenamefont
		{Frisk~Kockum}, \citenamefont {Delsing},\ and\ \citenamefont
		{Johansson}}]{Kockum:14}%
	\BibitemOpen
	\bibfield  {author} {\bibinfo {author} {\bibfnamefont {A.}~\bibnamefont
			{Frisk~Kockum}}, \bibinfo {author} {\bibfnamefont {P.}~\bibnamefont
			{Delsing}},  and \bibinfo {author} {\bibfnamefont {G.}~\bibnamefont
			{Johansson}},\ }\enquote{\bibinfo {title} {Designing frequency-dependent
			relaxation rates and Lamb shifts for a giant artificial atom}},\ \href
	{\doibase 10.1103/PhysRevA.90.013837} {\bibfield  {journal} {\bibinfo
			{journal} {Phys. Rev. A}\ }\textbf {\bibinfo {volume} {90}},\ \bibinfo
		{pages} {013837} (\bibinfo {year} {2014})}\BibitemShut {NoStop}%
	\bibitem [{\citenamefont {Kockum}(2020)}]{Kockum:20}%
	\BibitemOpen
	\bibfield  {author} {\bibinfo {author} {\bibfnamefont {A.~F.}\ \bibnamefont
			{Kockum}},\ }in\ \href@noop {} {\emph {\bibinfo {booktitle} {International
				Symposium on Mathematics, Quantum Theory, and Cryptography}}}\ (\bibinfo
	{organization} {Springer, Singapore},\ \bibinfo {year} {2020})\ pp.\ \bibinfo
	{pages} {125--146}\BibitemShut {NoStop}%
	\bibitem [{\citenamefont {Manenti}\ \emph {et~al.}(2017)\citenamefont
		{Manenti}, \citenamefont {Kockum}, \citenamefont {Patterson}, \citenamefont
		{Behrle}, \citenamefont {Rahamim}, \citenamefont {Tancredi}, \citenamefont
		{Nori},\ and\ \citenamefont {Leek}}]{Manenti:17}%
	\BibitemOpen
	\bibfield  {author} {\bibinfo {author} {\bibfnamefont {R.}~\bibnamefont
			{Manenti}},  {\em et~al.},\ }\enquote{\bibinfo {title} {Circuit quantum
			acoustodynamics with surface acoustic waves Nat}},\ \href {\doibase
		10.1038/s41467-017-01063-9} {\bibfield  {journal} {\bibinfo  {journal}
			{Nature Communications}\ }\textbf {\bibinfo {volume} {8}},\ \bibinfo {pages}
		{975} (\bibinfo {year} {2017})}\BibitemShut {NoStop}%
	\bibitem [{\citenamefont {Noguchi}\ \emph {et~al.}(2017)\citenamefont
		{Noguchi}, \citenamefont {Yamazaki}, \citenamefont {Tabuchi},\ and\
		\citenamefont {Nakamura}}]{Noguchi:17}%
	\BibitemOpen
	\bibfield  {author} {\bibinfo {author} {\bibfnamefont {A.}~\bibnamefont
			{Noguchi}}, \bibinfo {author} {\bibfnamefont {R.}~\bibnamefont {Yamazaki}},
		\bibinfo {author} {\bibfnamefont {Y.}~\bibnamefont {Tabuchi}},  and \bibinfo
		{author} {\bibfnamefont {Y.}~\bibnamefont {Nakamura}},\ }\enquote{\bibinfo
		{title} {Qubit-Assisted Transduction for a Detection of Surface Acoustic
			Waves near the Quantum Limit}},\ \href {\doibase
		10.1103/PhysRevLett.119.180505} {\bibfield  {journal} {\bibinfo  {journal}
			{Phys. Rev. Lett.}\ }\textbf {\bibinfo {volume} {119}},\ \bibinfo {pages}
		{180505} (\bibinfo {year} {2017})}\BibitemShut {NoStop}%
	\bibitem [{\citenamefont {Bolgar}\ \emph {et~al.}(2018)\citenamefont {Bolgar},
		\citenamefont {Zotova}, \citenamefont {Kirichenko}, \citenamefont {Besedin},
		\citenamefont {Semenov}, \citenamefont {Shaikhaidarov},\ and\ \citenamefont
		{Astafiev}}]{Bolgar:18}%
	\BibitemOpen
	\bibfield  {author} {\bibinfo {author} {\bibfnamefont {A.~N.}\ \bibnamefont
			{Bolgar}},  {\em et~al.},\ }\enquote{\bibinfo {title} {Quantum Regime of a
			Two-Dimensional Phonon Cavity}},\ \href {\doibase
		10.1103/PhysRevLett.120.223603} {\bibfield  {journal} {\bibinfo  {journal}
			{Phys. Rev. Lett.}\ }\textbf {\bibinfo {volume} {120}},\ \bibinfo {pages}
		{223603} (\bibinfo {year} {2018})}\BibitemShut {NoStop}%
	\bibitem [{\citenamefont {Vadiraj}\ \emph {et~al.}(2021)\citenamefont
		{Vadiraj}, \citenamefont {Ask}, \citenamefont {McConkey}, \citenamefont
		{Nsanzineza}, \citenamefont {Chang}, \citenamefont {Kockum},\ and\
		\citenamefont {Wilson}}]{Vadiraj:21}%
	\BibitemOpen
	\bibfield  {author} {\bibinfo {author} {\bibfnamefont {A.~M.}\ \bibnamefont
			{Vadiraj}},  {\em et~al.},\ }\enquote{\bibinfo {title} {Engineering the level
			structure of a giant artificial atom in waveguide quantum electrodynamics}},\
	\href {\doibase 10.1103/PhysRevA.103.023710} {\bibfield  {journal} {\bibinfo
			{journal} {Phys. Rev. A}\ }\textbf {\bibinfo {volume} {103}},\ \bibinfo
		{pages} {023710} (\bibinfo {year} {2021})}\BibitemShut {NoStop}%
	\bibitem [{\citenamefont {Wang}\ \emph {et~al.}(2021)\citenamefont {Wang},
		\citenamefont {Liu}, \citenamefont {Kockum}, \citenamefont {Li},\ and\
		\citenamefont {Nori}}]{Wang:21}%
	\BibitemOpen
	\bibfield  {author} {\bibinfo {author} {\bibfnamefont {X.}~\bibnamefont
			{Wang}},  {\em et~al.},\ }\enquote{\bibinfo {title} {Tunable Chiral Bound
			States with Giant Atoms}},\ \href {\doibase 10.1103/PhysRevLett.126.043602}
	{\bibfield  {journal} {\bibinfo  {journal} {Phys. Rev. Lett.}\ }\textbf
		{\bibinfo {volume} {126}},\ \bibinfo {pages} {043602} (\bibinfo {year}
		{2021})}\BibitemShut {NoStop}%
	\bibitem [{\citenamefont {Du}\ \emph {et~al.}(2022{\natexlab{a}})\citenamefont
		{Du}, \citenamefont {Zhang}, \citenamefont {Wu}, \citenamefont {Kockum},\
		and\ \citenamefont {Li}}]{Du:22}%
	\BibitemOpen
	\bibfield  {author} {\bibinfo {author} {\bibfnamefont {L.}~\bibnamefont
			{Du}},  {\em et~al.},\ }\enquote{\bibinfo {title} {Giant Atoms in a Synthetic
			Frequency Dimension}},\ \href {\doibase 10.1103/PhysRevLett.128.223602}
	{\bibfield  {journal} {\bibinfo  {journal} {Phys. Rev. Lett.}\ }\textbf
		{\bibinfo {volume} {128}},\ \bibinfo {pages} {223602} (\bibinfo {year}
		{2022}{\natexlab{a}})}\BibitemShut {NoStop}%
	\bibitem [{\citenamefont {Zhao}\ and\ \citenamefont {Wang}(2020)}]{Zhao:20}%
	\BibitemOpen
	\bibfield  {author} {\bibinfo {author} {\bibfnamefont {W.}~\bibnamefont
			{Zhao}} and \bibinfo {author} {\bibfnamefont {Z.}~\bibnamefont {Wang}},\
	}\enquote{\bibinfo {title} {Single-photon scattering and bound states in an
			atom-waveguide system with two or multiple coupling points}},\ \href
	{\doibase 10.1103/PhysRevA.101.053855} {\bibfield  {journal} {\bibinfo
			{journal} {Phys. Rev. A}\ }\textbf {\bibinfo {volume} {101}},\ \bibinfo
		{pages} {053855} (\bibinfo {year} {2020})}\BibitemShut {NoStop}%
	\bibitem [{\citenamefont {Kockum}\ \emph {et~al.}(2018)\citenamefont {Kockum},
		\citenamefont {Johansson},\ and\ \citenamefont {Nori}}]{Kockum:18}%
	\BibitemOpen
	\bibfield  {author} {\bibinfo {author} {\bibfnamefont {A.~F.}\ \bibnamefont
			{Kockum}}, \bibinfo {author} {\bibfnamefont {G.}~\bibnamefont {Johansson}},
		and \bibinfo {author} {\bibfnamefont {F.}~\bibnamefont {Nori}},\
	}\enquote{\bibinfo {title} {Decoherence-Free Interaction between Giant Atoms
			in Waveguide Quantum Electrodynamics}},\ \href {\doibase
		10.1103/PhysRevLett.120.140404} {\bibfield  {journal} {\bibinfo  {journal}
			{Phys. Rev. Lett.}\ }\textbf {\bibinfo {volume} {120}},\ \bibinfo {pages}
		{140404} (\bibinfo {year} {2018})}\BibitemShut {NoStop}%
	\bibitem [{\citenamefont {Soro}\ and\ \citenamefont {Kockum}(2022)}]{Soro:22}%
	\BibitemOpen
	\bibfield  {author} {\bibinfo {author} {\bibfnamefont {A.}~\bibnamefont
			{Soro}} and \bibinfo {author} {\bibfnamefont {A.~F.}\ \bibnamefont
			{Kockum}},\ }\enquote{\bibinfo {title} {Chiral quantum optics with giant
			atoms}},\ \href {\doibase 10.1103/PhysRevA.105.023712} {\bibfield  {journal}
		{\bibinfo  {journal} {Phys. Rev. A}\ }\textbf {\bibinfo {volume} {105}},\
		\bibinfo {pages} {023712} (\bibinfo {year} {2022})}\BibitemShut {NoStop}%
	\bibitem [{\citenamefont {Kannan}\ \emph {et~al.}(2020)\citenamefont {Kannan},
		\citenamefont {Ruckriegel}, \citenamefont {Campbell}, \citenamefont
		{Frisk~Kockum}, \citenamefont {Braumüller}, \citenamefont {Kim},
		\citenamefont {Kjaergaard}, \citenamefont {Krantz}, \citenamefont {Melville},
		\citenamefont {Niedzielski}, \citenamefont {Vepsäläinen}, \citenamefont
		{Winik}, \citenamefont {Yoder}, \citenamefont {Nori}, \citenamefont
		{Orlando}, \citenamefont {Gustavsson},\ and\ \citenamefont
		{Oliver}}]{Kannan:20}%
	\BibitemOpen
	\bibfield  {author} {\bibinfo {author} {\bibfnamefont {B.}~\bibnamefont
			{Kannan}},  {\em et~al.},\ }\enquote{\bibinfo {title} {Waveguide quantum
			electrodynamics with superconducting artificial giant atoms}},\ \href
	{\doibase 10.1038/s41586-020-2529-9} {\bibfield  {journal} {\bibinfo
			{journal} {Nature}\ }\textbf {\bibinfo {volume} {583}},\ \bibinfo {pages}
		{775} (\bibinfo {year} {2020})}\BibitemShut {NoStop}%
	\bibitem [{\citenamefont {Yu}\ \emph {et~al.}(2021)\citenamefont {Yu},
		\citenamefont {Wang},\ and\ \citenamefont {Wu}}]{Yu:21}%
	\BibitemOpen
	\bibfield  {author} {\bibinfo {author} {\bibfnamefont {H.}~\bibnamefont
			{Yu}}, \bibinfo {author} {\bibfnamefont {Z.}~\bibnamefont {Wang}},  and
		\bibinfo {author} {\bibfnamefont {J.-H.}\ \bibnamefont {Wu}},\
	}\enquote{\bibinfo {title} {Entanglement preparation and nonreciprocal
			excitation evolution in giant atoms by controllable dissipation and
			coupling}},\ \href {\doibase 10.1103/PhysRevA.104.013720} {\bibfield
		{journal} {\bibinfo  {journal} {Phys. Rev. A}\ }\textbf {\bibinfo {volume}
			{104}},\ \bibinfo {pages} {013720} (\bibinfo {year} {2021})}\BibitemShut
	{NoStop}%
	\bibitem [{\citenamefont {Barends}\ \emph {et~al.}(2014)\citenamefont
		{Barends}, \citenamefont {Kelly}, \citenamefont {Megrant}, \citenamefont
		{Veitia}, \citenamefont {Sank}, \citenamefont {Jeffrey}, \citenamefont
		{White}, \citenamefont {Mutus}, \citenamefont {Fowler}, \citenamefont
		{Campbell}, \citenamefont {Chen}, \citenamefont {Chen}, \citenamefont
		{Chiaro}, \citenamefont {Dunsworth}, \citenamefont {Neill}, \citenamefont
		{O’Malley}, \citenamefont {Roushan}, \citenamefont {Vainsencher},
		\citenamefont {Wenner}, \citenamefont {Korotkov}, \citenamefont {Cleland},\
		and\ \citenamefont {Martinis}}]{Barends:14}%
	\BibitemOpen
	\bibfield  {author} {\bibinfo {author} {\bibfnamefont {R.}~\bibnamefont
			{Barends}},  {\em et~al.},\ }\enquote{\bibinfo {title} {Superconducting
			quantum circuits at the surface code threshold for fault tolerance}},\ \href
	{\doibase 10.1038/nature13171} {\bibfield  {journal} {\bibinfo  {journal}
			{Nature}\ }\textbf {\bibinfo {volume} {508}},\ \bibinfo {pages} {500}
		(\bibinfo {year} {2014})}\BibitemShut {NoStop}%
	\bibitem [{\citenamefont {Krantz}\ \emph {et~al.}(2019)\citenamefont {Krantz},
		\citenamefont {Kjaergaard}, \citenamefont {Yan}, \citenamefont {Orlando},
		\citenamefont {Gustavsson},\ and\ \citenamefont {Oliver}}]{krantz:19}%
	\BibitemOpen
	\bibfield  {author} {\bibinfo {author} {\bibfnamefont {P.}~\bibnamefont
			{Krantz}},  {\em et~al.},\ }\enquote{\bibinfo {title} {A quantum engineer's
			guide to superconducting qubits}},\ \href {\doibase 10.1063/1.5089550}
	{\bibfield  {journal} {\bibinfo  {journal} {Applied Physics Reviews}\
		}\textbf {\bibinfo {volume} {6}},\ \bibinfo {pages} {021318} (\bibinfo {year}
		{2019})}\BibitemShut {NoStop}%
	\bibitem [{Sup()}]{SupInf_GiantAtoms}%
	\BibitemOpen
	\href@noop {} {\bibinfo  {journal} {See Supplementary Material [url] for more
			details about the master equation used to describe the system, for further
			information of the lifetime discussed in the paper, and for discussion
			super-radiant entanglement generation in small atoms, which includes
			Refs.~\cite{Combes:17,John:09,Peyman:19,Mandel:Book}}\ }\BibitemShut
	{NoStop}%
	\bibitem [{\citenamefont {Du}\ \emph {et~al.}(2022{\natexlab{b}})\citenamefont
		{Du}, \citenamefont {Chen}, \citenamefont {Zhang},\ and\ \citenamefont
		{Li}}]{Lei:22}%
	\BibitemOpen
	\bibfield  {journal} {  }\bibfield  {author} {\bibinfo {author} {\bibfnamefont
			{L.}~\bibnamefont {Du}}, \bibinfo {author} {\bibfnamefont {Y.-T.}\
			\bibnamefont {Chen}}, \bibinfo {author} {\bibfnamefont {Y.}~\bibnamefont
			{Zhang}},  and \bibinfo {author} {\bibfnamefont {Y.}~\bibnamefont {Li}},\
	}\enquote{\bibinfo {title} {Giant atoms with time-dependent couplings}},\
	\href {\doibase 10.1103/PhysRevResearch.4.023198} {\bibfield  {journal}
		{\bibinfo  {journal} {Phys. Rev. Research}\ }\textbf {\bibinfo {volume}
			{4}},\ \bibinfo {pages} {023198} (\bibinfo {year}
		{2022}{\natexlab{b}})}\BibitemShut {NoStop}%
	\bibitem [{\citenamefont {Wang}\ and\ \citenamefont {Li}(2022)}]{Wang:22}%
	\BibitemOpen
	\bibfield  {author} {\bibinfo {author} {\bibfnamefont {X.}~\bibnamefont
			{Wang}} and \bibinfo {author} {\bibfnamefont {H.-R.}\ \bibnamefont {Li}},\
	}\enquote{\bibinfo {title} {Chiral quantum network with giant atoms}},\ \href
	{\doibase 10.1088/2058-9565/ac6a04} {\bibfield  {journal} {\bibinfo
			{journal} {Quantum Science and Technology}\ }\textbf {\bibinfo {volume}
			{7}},\ \bibinfo {pages} {035007} (\bibinfo {year} {2022})}\BibitemShut
	{NoStop}%
	\bibitem [{\citenamefont {{Du}}\ \emph {et~al.}(2022)\citenamefont {{Du}},
		\citenamefont {{Zhang}},\ and\ \citenamefont {{Li}}}]{Lei:22b}%
	\BibitemOpen
	\bibfield  {author} {\bibinfo {author} {\bibfnamefont {L.}~\bibnamefont
			{{Du}}}, \bibinfo {author} {\bibfnamefont {Y.}~\bibnamefont {{Zhang}}},  and
		\bibinfo {author} {\bibfnamefont {Y.}~\bibnamefont {{Li}}},\
	}\enquote{\bibinfo {title} {{Giant atoms with modulated transition
				frequency}}},\ \href@noop {} {\bibfield  {journal} {\bibinfo  {journal}
			{arXiv e-prints}\ ,\ \bibinfo {eid} {arXiv:2206.14974}} (\bibinfo {year}
		{2022})},\ \Eprint {http://arxiv.org/abs/2206.14974} {arXiv:2206.14974
		[quant-ph]} \BibitemShut {NoStop}%
	\bibitem [{\citenamefont {{Qiu}}\ \emph {et~al.}(2022)\citenamefont {{Qiu}},
		\citenamefont {{Wu}},\ and\ \citenamefont {{L{\"u}}}}]{Qiu:22}%
	\BibitemOpen
	\bibfield  {author} {\bibinfo {author} {\bibfnamefont {Q.-Y.}\ \bibnamefont
			{{Qiu}}}, \bibinfo {author} {\bibfnamefont {Y.}~\bibnamefont {{Wu}}},  and
		\bibinfo {author} {\bibfnamefont {X.-Y.}\ \bibnamefont {{L{\"u}}}},\
	}\enquote{\bibinfo {title} {{Collective Radiance of Giant Atoms in
				Non-Markovian Regime}}},\ \href@noop {} {\bibfield  {journal} {\bibinfo
			{journal} {arXiv e-prints}\ ,\ \bibinfo {eid} {arXiv:2205.10982}} (\bibinfo
		{year} {2022})},\ \Eprint {http://arxiv.org/abs/2205.10982} {arXiv:2205.10982
		[quant-ph]} \BibitemShut {NoStop}%
	\bibitem [{\citenamefont {Yin}\ \emph {et~al.}(2022)\citenamefont {Yin},
		\citenamefont {Luo},\ and\ \citenamefont {Liao}}]{Yin:23}%
	\BibitemOpen
	\bibfield  {author} {\bibinfo {author} {\bibfnamefont {X.-L.}\ \bibnamefont
			{Yin}}, \bibinfo {author} {\bibfnamefont {W.-B.}\ \bibnamefont {Luo}},  and
		\bibinfo {author} {\bibfnamefont {J.-Q.}\ \bibnamefont {Liao}},\
	}\enquote{\bibinfo {title} {Non-Markovian disentanglement dynamics in
			double-giant-atom waveguide-QED systems}},\ \href {\doibase
		10.1103/PhysRevA.106.063703} {\bibfield  {journal} {\bibinfo  {journal}
			{Phys. Rev. A}\ }\textbf {\bibinfo {volume} {106}},\ \bibinfo {pages}
		{063703} (\bibinfo {year} {2022})}\BibitemShut {NoStop}%
	\bibitem [{\citenamefont {Dicke}(1954)}]{Dicke:54}%
	\BibitemOpen
	\bibfield  {author} {\bibinfo {author} {\bibfnamefont {R.~H.}\ \bibnamefont
			{Dicke}},\ }\enquote{\bibinfo {title} {Coherence in Spontaneous Radiation
			Processes}},\ \href {\doibase 10.1103/PhysRev.93.99} {\bibfield  {journal}
		{\bibinfo  {journal} {Phys. Rev.}\ }\textbf {\bibinfo {volume} {93}},\
		\bibinfo {pages} {99} (\bibinfo {year} {1954})}\BibitemShut {NoStop}%
	\bibitem [{\citenamefont {Tana}\ and\ \citenamefont
		{Ficek}(2004)}]{Tanas-Ficek:04}%
	\BibitemOpen
	\bibfield  {author} {\bibinfo {author} {\bibfnamefont {R.}~\bibnamefont
			{Tana}} and \bibinfo {author} {\bibfnamefont {Z.}~\bibnamefont {Ficek}},\
	}\enquote{\bibinfo {title} {Entangling two atoms via spontaneous emission}},\
	\href {\doibase 10.1088/1464-4266/6/3/015} {\bibfield  {journal} {\bibinfo
			{journal} {Journal of Optics B: Quantum and Semiclassical Optics}\ }\textbf
		{\bibinfo {volume} {6}},\ \bibinfo {pages} {S90} (\bibinfo {year}
		{2004})}\BibitemShut {NoStop}%
	\bibitem [{\citenamefont {Cidrim}\ \emph {et~al.}(2020)\citenamefont {Cidrim},
		\citenamefont {do~Espirito~Santo}, \citenamefont {Schachenmayer},
		\citenamefont {Kaiser},\ and\ \citenamefont {Bachelard}}]{Cidrim:20}%
	\BibitemOpen
	\bibfield  {author} {\bibinfo {author} {\bibfnamefont {A.}~\bibnamefont
			{Cidrim}},  {\em et~al.},\ }\enquote{\bibinfo {title} {Photon Blockade with
			Ground-State Neutral Atoms}},\ \href {\doibase
		10.1103/PhysRevLett.125.073601} {\bibfield  {journal} {\bibinfo  {journal}
			{Phys. Rev. Lett.}\ }\textbf {\bibinfo {volume} {125}},\ \bibinfo {pages}
		{073601} (\bibinfo {year} {2020})}\BibitemShut {NoStop}%
	\bibitem [{\citenamefont {Williamson}\ \emph {et~al.}(2020)\citenamefont
		{Williamson}, \citenamefont {Borgh},\ and\ \citenamefont
		{Ruostekoski}}]{Williamson:20}%
	\BibitemOpen
	\bibfield  {author} {\bibinfo {author} {\bibfnamefont {L.~A.}\ \bibnamefont
			{Williamson}}, \bibinfo {author} {\bibfnamefont {M.~O.}\ \bibnamefont
			{Borgh}},  and \bibinfo {author} {\bibfnamefont {J.}~\bibnamefont
			{Ruostekoski}},\ }\enquote{\bibinfo {title} {Superatom Picture of Collective
			Nonclassical Light Emission and Dipole Blockade in Atom Arrays}},\ \href
	{\doibase 10.1103/PhysRevLett.125.073602} {\bibfield  {journal} {\bibinfo
			{journal} {Phys. Rev. Lett.}\ }\textbf {\bibinfo {volume} {125}},\ \bibinfo
		{pages} {073602} (\bibinfo {year} {2020})}\BibitemShut {NoStop}%
	\bibitem [{\citenamefont {Trebbia}\ \emph {et~al.}(2022)\citenamefont
		{Trebbia}, \citenamefont {Deplano}, \citenamefont {Tamarat},\ and\
		\citenamefont {Lounis}}]{Trebbia:22}%
	\BibitemOpen
	\bibfield  {author} {\bibinfo {author} {\bibfnamefont {J.-B.}\ \bibnamefont
			{Trebbia}}, \bibinfo {author} {\bibfnamefont {Q.}~\bibnamefont {Deplano}},
		\bibinfo {author} {\bibfnamefont {P.}~\bibnamefont {Tamarat}},  and \bibinfo
		{author} {\bibfnamefont {B.}~\bibnamefont {Lounis}},\ }\enquote{\bibinfo
		{title} {Tailoring the superradiant and subradiant nature of two coherently
			coupled quantum emitters}},\ \href {\doibase 10.1038/s41467-022-30672-2}
	{\bibfield  {journal} {\bibinfo  {journal} {Nature communications}\ }\textbf
		{\bibinfo {volume} {13}},\ \bibinfo {pages} {1} (\bibinfo {year}
		{2022})}\BibitemShut {NoStop}%
	\bibitem [{Note1()}]{Note1}%
	\BibitemOpen
	\bibinfo {note} {Note that this state can be written as a linear combination
		of the sub- and sub-radiant modes $\ket {\psi _{\pm }}$ introduced before for
		giant atoms.}\BibitemShut {Stop}%
	\bibitem [{\citenamefont {Hill}\ and\ \citenamefont
		{Wootters}(1997)}]{Hill:97}%
	\BibitemOpen
	\bibfield  {author} {\bibinfo {author} {\bibfnamefont {S.}~\bibnamefont
			{Hill}} and \bibinfo {author} {\bibfnamefont {W.~K.}\ \bibnamefont
			{Wootters}},\ }\enquote{\bibinfo {title} {Entanglement of a Pair of Quantum
			Bits}},\ \href {\doibase 10.1103/PhysRevLett.78.5022} {\bibfield  {journal}
		{\bibinfo  {journal} {Phys. Rev. Lett.}\ }\textbf {\bibinfo {volume} {78}},\
		\bibinfo {pages} {5022} (\bibinfo {year} {1997})}\BibitemShut {NoStop}%
	\bibitem [{\citenamefont {Nielsen}\ and\ \citenamefont
		{Chuang}(2011)}]{Nielsen:Book}%
	\BibitemOpen
	\bibfield  {author} {\bibinfo {author} {\bibfnamefont {M.~A.}\ \bibnamefont
			{Nielsen}} and \bibinfo {author} {\bibfnamefont {I.~L.}\ \bibnamefont
			{Chuang}},\ }\href {\doibase 10.1017/CBO9780511976667} {\emph {\bibinfo
			{title} {Quantum Computation and Quantum Information: 10th Anniversary
				Edition}}},\ \bibinfo {edition} {10th}\ ed.\ (\bibinfo  {publisher}
	{Cambridge University Press},\ \bibinfo {address} {New York, NY, USA},\
	\bibinfo {year} {2011})\BibitemShut {NoStop}%
	\bibitem [{\citenamefont {Mandel}(1979)}]{Mandel:79}%
	\BibitemOpen
	\bibfield  {author} {\bibinfo {author} {\bibfnamefont {L.}~\bibnamefont
			{Mandel}},\ }\enquote{\bibinfo {title} {Sub-Poissonian photon statistics in
			resonance fluorescence}},\ \href {\doibase 10.1364/OL.4.000205} {\bibfield
		{journal} {\bibinfo  {journal} {Opt. Lett.}\ }\textbf {\bibinfo {volume}
			{4}},\ \bibinfo {pages} {205} (\bibinfo {year} {1979})}\BibitemShut {NoStop}%
	\bibitem [{\citenamefont {Jahnke}\ \emph {et~al.}(2016)\citenamefont {Jahnke},
		\citenamefont {Gies}, \citenamefont {A{\ss}mann}, \citenamefont {Bayer},
		\citenamefont {Leymann}, \citenamefont {Foerster}, \citenamefont {Wiersig},
		\citenamefont {Schneider}, \citenamefont {Kamp},\ and\ \citenamefont
		{H{\"o}fling}}]{Jahnke:16}%
	\BibitemOpen
	\bibfield  {author} {\bibinfo {author} {\bibfnamefont {F.}~\bibnamefont
			{Jahnke}},  {\em et~al.},\ }\enquote{\bibinfo {title} {Giant photon bunching,
			superradiant pulse emission and excitation trapping in quantum-dot
			nanolasers}},\ \href {\doibase https://doi.org/10.1038/ncomms11540}
	{\bibfield  {journal} {\bibinfo  {journal} {Nature communications}\ }\textbf
		{\bibinfo {volume} {7}},\ \bibinfo {pages} {1} (\bibinfo {year}
		{2016})}\BibitemShut {NoStop}%
	\bibitem [{\citenamefont {Arcari}\ \emph {et~al.}(2014)\citenamefont {Arcari},
		\citenamefont {S\"ollner}, \citenamefont {Javadi}, \citenamefont
		{Lindskov~Hansen}, \citenamefont {Mahmoodian}, \citenamefont {Liu},
		\citenamefont {Thyrrestrup}, \citenamefont {Lee}, \citenamefont {Song},
		\citenamefont {Stobbe},\ and\ \citenamefont {Lodahl}}]{Arcari:14}%
	\BibitemOpen
	\bibfield  {author} {\bibinfo {author} {\bibfnamefont {M.}~\bibnamefont
			{Arcari}},  {\em et~al.},\ }\enquote{\bibinfo {title} {Near-Unity Coupling
			Efficiency of a Quantum Emitter to a Photonic Crystal Waveguide}},\ \href
	{\doibase 10.1103/PhysRevLett.113.093603} {\bibfield  {journal} {\bibinfo
			{journal} {Phys. Rev. Lett.}\ }\textbf {\bibinfo {volume} {113}},\ \bibinfo
		{pages} {093603} (\bibinfo {year} {2014})}\BibitemShut {NoStop}%
	\bibitem [{\citenamefont {Gin\'es}\ \emph {et~al.}(2022)\citenamefont
		{Gin\'es}, \citenamefont {Mocza\l{}a-Dusanowska}, \citenamefont {Dlaka},
		\citenamefont {Ho\ifmmode~\check{s}\else \v{s}\fi{}\'ak}, \citenamefont
		{Gonzales-Ureta}, \citenamefont {Lee}, \citenamefont
		{Je\ifmmode~\check{z}\else \v{z}\fi{}ek}, \citenamefont {Harbord},
		\citenamefont {Oulton}, \citenamefont {H\"ofling}, \citenamefont {Young},
		\citenamefont {Schneider},\ and\ \citenamefont
		{Predojevi\ifmmode~\acute{c}\else \'{c}\fi{}}}]{Gines:22}%
	\BibitemOpen
	\bibfield  {author} {\bibinfo {author} {\bibfnamefont {L.}~\bibnamefont
			{Gin\'es}},  {\em et~al.},\ }\enquote{\bibinfo {title} {High Extraction
			Efficiency Source of Photon Pairs Based on a Quantum Dot Embedded in a
			Broadband Micropillar Cavity}},\ \href {\doibase
		10.1103/PhysRevLett.129.033601} {\bibfield  {journal} {\bibinfo  {journal}
			{Phys. Rev. Lett.}\ }\textbf {\bibinfo {volume} {129}},\ \bibinfo {pages}
		{033601} (\bibinfo {year} {2022})}\BibitemShut {NoStop}%
	\bibitem [{\citenamefont {Combes}\ \emph {et~al.}(2017)\citenamefont {Combes},
		\citenamefont {Kerckhoff},\ and\ \citenamefont {Sarovar}}]{Combes:17}%
	\BibitemOpen
	\bibfield  {author} {\bibinfo {author} {\bibfnamefont {J.}~\bibnamefont
			{Combes}}, \bibinfo {author} {\bibfnamefont {J.}~\bibnamefont {Kerckhoff}},
		and \bibinfo {author} {\bibfnamefont {M.}~\bibnamefont {Sarovar}},\
	}\enquote{\bibinfo {title} {The SLH framework for modeling quantum
			input-output networks}},\ \href {\doibase 10.1080/23746149.2017.1343097}
	{\bibfield  {journal} {\bibinfo  {journal} {Advances in Physics: X}\ }\textbf
		{\bibinfo {volume} {2}},\ \bibinfo {pages} {784} (\bibinfo {year}
		{2017})}\BibitemShut {NoStop}%
	\bibitem [{\citenamefont {Gough}\ and\ \citenamefont {James}(2009)}]{John:09}%
	\BibitemOpen
	\bibfield  {author} {\bibinfo {author} {\bibfnamefont {J.}~\bibnamefont
			{Gough}} and \bibinfo {author} {\bibfnamefont {M.~R.}\ \bibnamefont
			{James}},\ }\enquote{\bibinfo {title} {The Series Product and Its Application
			to Quantum Feedforward and Feedback Networks}},\ \href {\doibase
		10.1109/TAC.2009.2031205} {\bibfield  {journal} {\bibinfo  {journal} {IEEE
				Transactions on Automatic Control}\ }\textbf {\bibinfo {volume} {54}},\
		\bibinfo {pages} {2530} (\bibinfo {year} {2009})}\BibitemShut {NoStop}%
	\bibitem [{\citenamefont {Azodi}\ \emph {et~al.}(2019)\citenamefont {Azodi},
		\citenamefont {Setoodeh}, \citenamefont {Khayatian},\ and\ \citenamefont
		{Asemani}}]{Peyman:19}%
	\BibitemOpen
	\bibfield  {author} {\bibinfo {author} {\bibfnamefont {P.}~\bibnamefont
			{Azodi}}, \bibinfo {author} {\bibfnamefont {P.}~\bibnamefont {Setoodeh}},
		\bibinfo {author} {\bibfnamefont {A.}~\bibnamefont {Khayatian}},  and
		\bibinfo {author} {\bibfnamefont {M.~H.}\ \bibnamefont {Asemani}},\
	}\enquote{\bibinfo {title} {Uncertainty decomposition of quantum networks in
			SLH framework}},\ \href {\doibase https://doi.org/10.1002/rnc.4740}
	{\bibfield  {journal} {\bibinfo  {journal} {International Journal of Robust
				and Nonlinear Control}\ }\textbf {\bibinfo {volume} {29}},\ \bibinfo {pages}
		{6542} (\bibinfo {year} {2019})}\BibitemShut {NoStop}%
	\bibitem [{\citenamefont {Mandel}\ and\ \citenamefont
		{Wolf}(1995)}]{Mandel:Book}%
	\BibitemOpen
	\bibfield  {author} {\bibinfo {author} {\bibfnamefont {L.}~\bibnamefont
			{Mandel}} and \bibinfo {author} {\bibfnamefont {E.}~\bibnamefont {Wolf}},\
	}\href {\doibase 10.1017/CBO9781139644105} {\emph {\bibinfo {title} {Optical
				Coherence and Quantum Optics}}}\ (\bibinfo  {publisher} {Cambridge University
		Press},\ \bibinfo {year} {1995})\BibitemShut {NoStop}%
\end{thebibliography}

%

\onecolumngrid
\newpage

\begin{center}
	{\large{ {\bf Supplemental Material for: \\ Generation of maximally-entangled long-lived states with giant atoms in a waveguide}}}

\vskip0.5\baselineskip{Alan C. Santos~\orcidlink{0000-0002-6989-7958}$^{1,2,\ast}$, and R. Bachelard~\orcidlink{0000-0002-6026-509X}$^{1,3,\dagger}$}

\vskip0.5\baselineskip{{\em$^{1}$Departamento de Física, Universidade Federal de São Carlos,\\ Rodovia Washington Luís, km 235 - SP-310, 13565-905 São Carlos, SP, Brazil}\\
	{\em $^{2}$Department of Physics, Stockholm University, AlbaNova University Center 106 91 Stockholm, Sweden}
	\\
	{\em $^{3}$Universit\'e C\^ote d'Azur, CNRS, Institut de Physique de Nice, 06560 Valbonne, France}
}

\vskip0.5\baselineskip{$^{\color{blue}\ast}$ac\_santos@df.ufscar.br, ~~~ $^{\color{blue}\dagger}$romain@ufscar.br}
\end{center}

\appendix

\setcounter{equation}{0}
\setcounter{figure}{0}
\setcounter{table}{0}

\renewcommand{\theequation}{S\arabic{equation}}
\renewcommand{\thefigure}{S\arabic{figure}}

\twocolumngrid

	\section{Derivation of the master equation}

\begin{figure}[t!]
	\includegraphics[width=\columnwidth]{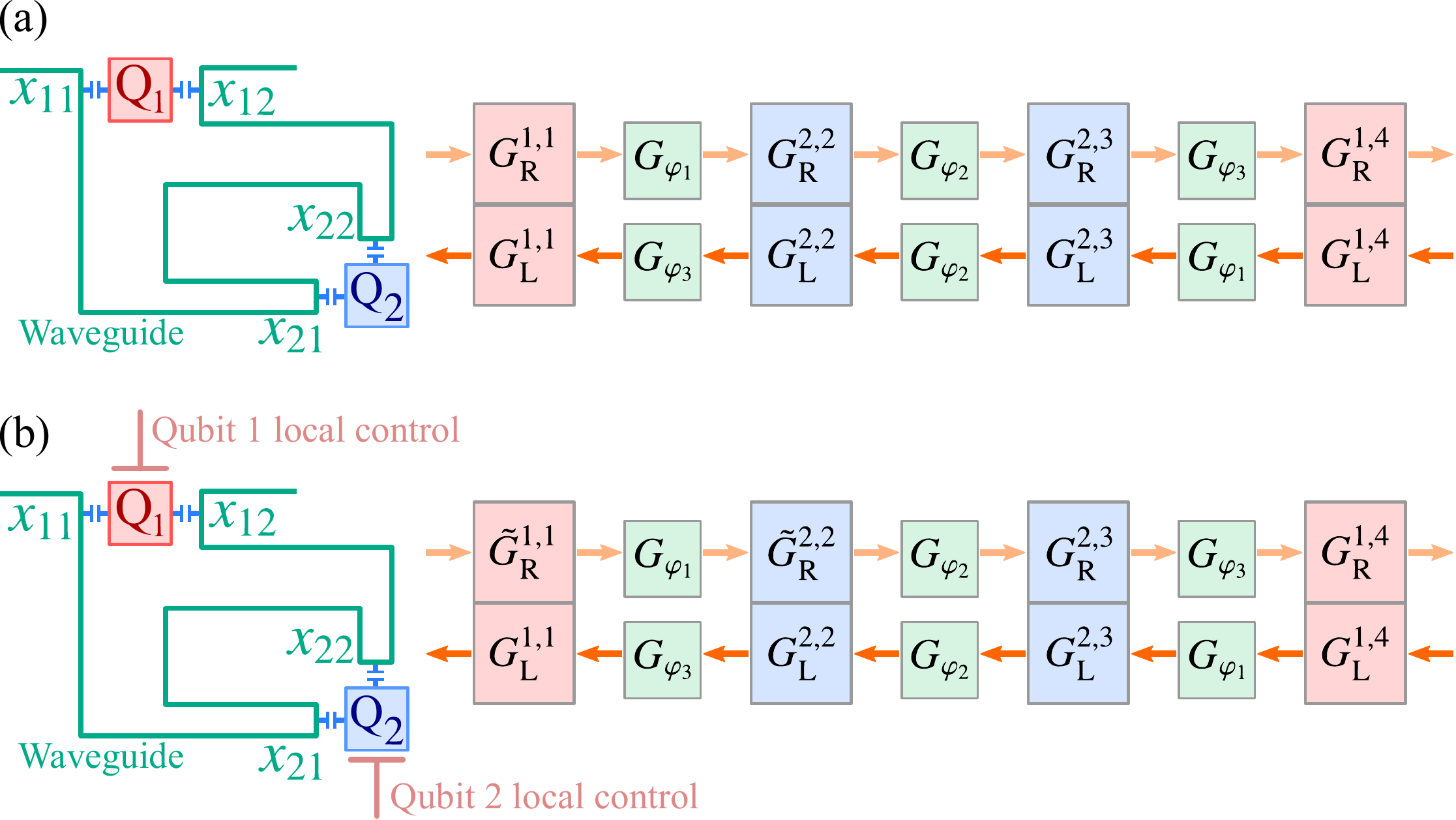}
	\caption{(a) Sketch of two nested giant superconducting atoms, where the atom 1 and 2 connect to waveguide, though a capacitive coupling, at points $(x_{11},x_{12})$ and $(x_{21},x_{22})$. The SLH diagram used to find the master equation~\cite{Kockum:18} is also shown. (b) Our schematic representation, where the local control can be done independently from the waveguide by capacitively coupling each atom to its own XY-drive line. In this way, the SLH diagram changes a bit according to the scheme shown in (b).}
	\label{SupMat-MasterEquation}
\end{figure}

In this section we present a brief derivation of the master equation which describes the giant atom dynamics when driven by an external field. First of all, we shortly describe how to obtain the master equation for the case where no external pump is applied to the atoms (see~\cite{Kockum:18} for a detailed discussion). Inspired by the experimental realization in superconducting atoms~\cite{Kannan:20}, Fig.~\ref{SupMat-MasterEquation}{\color{blue}a} shows the schematic representation of two nested giant atoms in superconducting circuit. Two artificial atoms (transmon qubits) are capacitively connected to a common waveguide. The master equation for this system is
\begin{align}
	\frac{d}{dt}\hat{\rho}(t) = \frac{1}{i\hbar}[ \hat{H}_{1}^{0} + \hat{H}_{2}^{0} , \hat{\rho}(t) ] + \Dcal (\hat{\rho}) , \label{ApEq:MasterEq}
\end{align}
where we define the Hamiltonian $\hat{H}_{n}^{0} = \hbar (\omega_{n} + \delta_{n})\hat\sigma^{+}_{n}\hat\sigma^{-}_{n}$ for each atom $n$, and $\Dcal (\hat{\rho})$ describes the effects of the interaction between the atoms and the waveguide (as defined in the main text). Now, we need to compute the term $\Dcal (\hat{\rho})$ to properly describe the dynamics of the system. To this end, we can use the SLH formalism~\cite{Combes:17}, which is also the  approach considered in Ref.~\cite{Kockum:18}. In this formalism we need the number $n$ of input and output ports to find the triplet of elements $\textbf{S}$ (a $n\times n$ scattering matrix), $\textbf{L}$ (a vector with $n$ matrices describing the system-environment interaction), and $\hat{H}$ the Hamiltonian of the system (including interactions~\cite{Combes:17}). The triplet is then denoted by $G = \left(\textbf{S}, \textbf{L},\hat{H} \right)$, and the system is completely described by the resulting triplet $G_{\mathrm{res}}$, obtained using SLH composition rules for $G_{k}$~\cite{Combes:17}. For example, for the case of two-nested atoms (with no pump) the SLH diagram is given by Fig.~\ref{SupMat-MasterEquation}{\color{blue}a}, and the resulting triplet is obtained from \textit{right}- and \textit{left}-moving photons through their \textit{independent} contributions encoded as
\begin{align}
	G_{\mathrm{R}} &= G_{\mathrm{R}}^{1,4} \triangleleft G_{\varphi_3} \triangleleft G_{\mathrm{R}}^{2,3} \triangleleft G_{\varphi_2} \triangleleft G_{\mathrm{R}}^{2,2} \triangleleft G_{\varphi_1} \triangleleft G_{\mathrm{R}}^{1, 1}, \\
	G_{\mathrm{L}} &= G_{\mathrm{L}}^{1,1} \triangleleft G_{\varphi_1} \triangleleft G_{\mathrm{L}}^{2,2} \triangleleft G_{\varphi_2} \triangleleft G_{\mathrm{L}}^{2,3} \triangleleft G_{\varphi_3} \triangleleft G_{\mathrm{L}}^{1, 4} ,
\end{align}
where we have defined
\begin{align}
	G_{\mathrm{R}}^{1, 1} &= \left(\1, \sqrt{\frac{\gamma_1}{2}}\hat{\sigma}_{1}^{-}, \hat{H}_{1}^{0} \right), ~
	G_{\mathrm{R}}^{1,4} = \left(\1, \sqrt{\frac{\gamma_4}{2}}\hat{\sigma}_{1}^{-}, 0 \right), \\
	G_{\mathrm{L}}^{1,1} &= \left(\1, \sqrt{\frac{\gamma_1}{2}}\hat{\sigma}_{1}^{-}, 0 \right), ~
	G_{\mathrm{L}}^{1,4} = \left(\1, \sqrt{\frac{\gamma_4}{2}}\hat{\sigma}_{1}^{-}, 0 \right), \\
	G_{\mathrm{R}}^{2,2} &= \left(\1, \sqrt{\frac{\gamma_2}{2}}\hat{\sigma}_{2}^{-},  \hat{H}_{2}^{0} \right), ~
	G_{\mathrm{R}}^{2,3} = \left(\1, \sqrt{\frac{\gamma_3}{2}}\hat{\sigma}_{2}^{-}, 0 \right), \\
	G_{\mathrm{L}}^{2,2} &= \left(\1, \sqrt{\frac{\gamma_2}{2}}\hat{\sigma}_{2}^{-}, 0 \right), ~
	G_{\mathrm{L}}^{2,3} = \left(\1, \sqrt{\frac{\gamma_3}{2}}\hat{\sigma}_{2}^{-}, 0 \right) ,
\end{align}
and the elements $G_{\varphi_n}$ are the phase acquired by the photon traveling into the waveguide. The notation $G_{n} \triangleleft G_{k}$ refers to the \textit{cascade rule} of the SLH formalism~\cite{Combes:17}. Therefore, the resulting triplet reads $G_{\mathrm{res}} = G_{\mathrm{R}} \boxplus G_{\mathrm{L}}$, in which $G_{n}\boxplus G_{k}$ denotes the \textit{concatenation product}.

Now, let us use the same methodology to include the local control of the atoms. In superconducting giant atoms the individual and local control of the atoms can be done by the direct capacitive coupling of each atom with their own XY-drive line~\cite{Barends:14,krantz:19}, as shown in Fig.~\ref{SupMat-MasterEquation}{\color{blue}b}. In this case, the Hamiltonian for the atoms, including control, is given by $\hat{H} = \hat{H}_{1} + \hat{H}_{2}$, with
\begin{align}
	\hat{H}_{n} = \hat{H}_{n}^{0} + \hat{H}_{n}^{\mathrm{c}} = \hat{H}_{n}^{0} + \hbar \Omega_{0}\left( \hat{\sigma}_{n}^{+} + \hat{\sigma}_{n}^{-}\right), 
\end{align}
with $\Omega_{0}$ the Rabi frequency of this pump. Here we highlight two important points: i) the above Hamiltonian does not depend on the driving field in the waveguide, and ii) the SLH rules are valid for any	$H$ on the localized system Hilbert space~\cite{Combes:17}. It means that the local external field can be suitably included by changing the local Hamiltonian $\hat{H}_{n}^{0}$ for another local Hamiltonian of the form $\hat{H}_{n}$. In fact, by using the SLH cascade rule given by~\cite{John:09,Peyman:19}
\begin{align}
	\left(\textbf{S}_{2}, \textbf{L}_{2}, \hat{H}_{2}  \right) \triangleleft \left( \textbf{S}_{1}, \textbf{L}_{1}, \hat{H}_{1} \right)   = \left(\textbf{S}_{2}\textbf{S}_{1}, \textbf{L}_{2} + \textbf{S}_{2}\textbf{L}_{1}, \hat{H}_{\mathrm{res}}  \right), \label{Ap:Eq:CascadeRule}
\end{align}
where the resulting Hamiltonian $\hat{H}_{\mathrm{res}}$ is given by
\begin{align}
	\hat{H}_{\mathrm{res}} = \hat{H}_{1} + \hat{H}_{2} + \mathrm{Im}\left( \textbf{L}^{\dagger}_{2} \textbf{S}_{2}\textbf{L}_{1}\right),
\end{align}
it is possible to conclude that
\begin{align}
	G_{\mathrm{R}}^{1, 1} \triangleleft \left(\1, \textbf{0}, \hat{H}_{1}^{\mathrm{c}} \right) = \left(\1, \sqrt{\frac{\gamma_1}{2}}\hat{\sigma}_{1}^{-}, \hat{H}_{1}^{0} + \hat{H}_{1}^{\mathrm{c}} \right) = \tilde{G}_{\mathrm{R}}^{1, 1}.
\end{align}

Therefore, by including an external local control $\hat{H}_{n}^{\mathrm{c}}$, we do not change the sector of the triplet $G_{\mathrm{R}}^{1, 1}$ associated to the dissipative term $\Dcal (\hat{\rho})$ and responsible for any interaction of the atoms with the waveguide. Only the Hamiltonian sector of $G_{\mathrm{R}}^{1, 1}$ is affected, but it does not promote any new interaction in the system. A similar calculation can be done to show that
\begin{align}
	G_{\mathrm{R}}^{2, 2} \triangleleft \left(\1, \textbf{0}, \hat{H}_{2}^{\mathrm{c}} \right) = \left(\1, \sqrt{\frac{\gamma_2}{2}}\hat{\sigma}_{2}^{-}, \hat{H}_{2}^{0} + \hat{H}_{2}^{\mathrm{c}} \right) = \tilde{G}_{\mathrm{R}}^{2, 2} ,
\end{align}
such that the new resulting SLH triplet is $\tilde{G}_{\mathrm{res}} = \tilde{G}_{\mathrm{R}} \boxplus G_{\mathrm{L}}$, where
\begin{align}
	\tilde{G}_{\mathrm{R}} &= G_{\mathrm{R}}^{1,4} \triangleleft G_{\varphi_3} \triangleleft G_{\mathrm{R}}^{2,3} \triangleleft G_{\varphi_2} \triangleleft \tilde{G}_{\mathrm{R}}^{2,2} \triangleleft G_{\varphi_1} \triangleleft \tilde{G}_{\mathrm{R}}^{1, 1} .
\end{align}

In Fig.~\ref{SupMat-MasterEquation}{\color{blue}b} we show the change promoted in the SLH diagram due to the local control on each atom.

\section{Computing decay rates}

Let us consider the two-atom system governed by the master equation given by
\begin{align}
	\frac{d}{dt}\hat{\rho}(t) = \Lcal[\hat{\rho} (t)] ,
\end{align}
where $\Lcal[\bullet]$ is a Louvillian that depends on the system we are dealing with. We consider a complete basis for the two-atom system, composed of the double excited state $\ket{ee}$, ground-state $\ket{gg}$, and $\ket{\psi_{\pm}}$ (defined in the main text). We are here interested in the decay from a state $\rho(0)$ to one of the target states $\{\ket{ee}, \ket{gg}, \ket{\psi_{\pm}}\}$. In this case, the evolved state can be computed as 
\begin{align}
	\hat{\rho}(t) &= p_g(t) \hat{\rho}_{gg} + p_e(t) \hat{\rho}_{ee} + p_{\psi_{+}}(t) \hat{\rho}_{\psi_{+}} + p_{\psi_{-}}(t) \hat{\rho}_{\psi_{-}} .
\end{align}
with $\hat{\rho}_{x} = \ket{x}\bra{x}$. Then, using these equations, the dynamics of each population in the above equation is
\begin{subequations}
	\begin{align}
		\dot{p}_{g}(t) &= p_g(t)\tr(\Lcal[\hat{\rho}_{gg}]\hat{\rho}_{gg}) + p_e(t)\tr(\Lcal[\hat{\rho}_{ee}]\hat{\rho}_{gg}) \nonumber \\
		&+  p_{\psi_{+}}(t)\tr(\Lcal[\hat{\rho}_{\psi_{+}}]\hat{\rho}_{gg}) + p_{\psi_{-}}(t)\tr(\Lcal[\hat{\rho}_{\psi_{-}}]\hat{\rho}_{gg}) , 
		\label{ApEq:p_g}\\
		\dot{p}_{e}(t) &= p_g(t)\tr(\Lcal[\hat{\rho}_{gg}]\hat{\rho}_{ee}) + p_e(t)\tr(\Lcal[\hat{\rho}_{ee}]\hat{\rho}_{ee}) \nonumber \\
		&+  p_{\psi_{+}}(t)\tr(\Lcal[\hat{\rho}_{\psi_{+}}]\hat{\rho}_{ee}) + p_{\psi_{-}}(t)\tr(\Lcal[\hat{\rho}_{\psi_{-}}]\hat{\rho}_{ee}) , 
		\\
		\dot{p}_{\psi_{\pm}}(t) &= p_g(t)\tr(\Lcal[\hat{\rho}_{gg}]\hat{\rho}_{\psi_{\pm}}) + p_e(t)\tr(\Lcal[\hat{\rho}_{ee}]\hat{\rho}_{\psi_{\pm}}) \nonumber \\
		&+  p_{\psi_{+}}(t)\tr(\Lcal[\hat{\rho}_{\psi_{+}}]\hat{\rho}_{\psi_{\pm}}) + p_{\psi_{-}}(t)\tr(\Lcal[\hat{\rho}_{\psi_{-}}]\hat{\rho}_{\psi_{\pm}}) .
	\end{align}
\end{subequations}

Let us discuss the meaning of these dynamical equations through the example of Eq.~\eqref{ApEq:p_g}. Now, to follow our claim, we need to take a particular form to $\Lcal[\bullet]$. For the master equation considered in the main text, defined in Eq.~({\color{blue}2}), it is possible to show that
\begin{align}
	\tr(\Lcal[\hat{\rho}_{gg}]\hat{\rho}_{gg}) = \tr(\Lcal[\hat{\rho}_{ee}]\hat{\rho}_{gg}) = 0 ,
\end{align}
such that
\begin{align}
	\dot{p}_{g}(t) &= p_{\psi_{+}}(t)\tr(\Lcal[\hat{\rho}_{\psi_{+}}]\hat{\rho}_{gg}) + p_{\psi_{-}}(t)\tr(\Lcal[\hat{\rho}_{\psi_{-}}]\hat{\rho}_{gg}) .
\end{align}

The gain of population in the ground state depends on the populations $ p_{\psi_{+}}(t)$ and $p_{\psi_{-}}(t)$, as well as the coefficients that set the decay rate of the states $\ket{\psi_{+}}$. Then, without loss of generality, we find the decay rate from the state $\ket{x}$ to $\ket{gg}$ in according with the above equation as
\begin{align}
	\Gamma_{x \rightarrow g} = \tr(\Lcal[\hat{\rho}_{x}]\hat{\rho}_{g}) .
\end{align}

We use this analysis to compute the decay rates shown in the main text. It is worth mentioning that the analysis done here is only valid for decay dynamics, as given by the master equation considered in the main text. Other kinds of decoherence, such as bit-flip or dephasing, for example, need to be studied on a case-by-case basis..

\begin{figure}[t!]
	\includegraphics[width=\columnwidth]{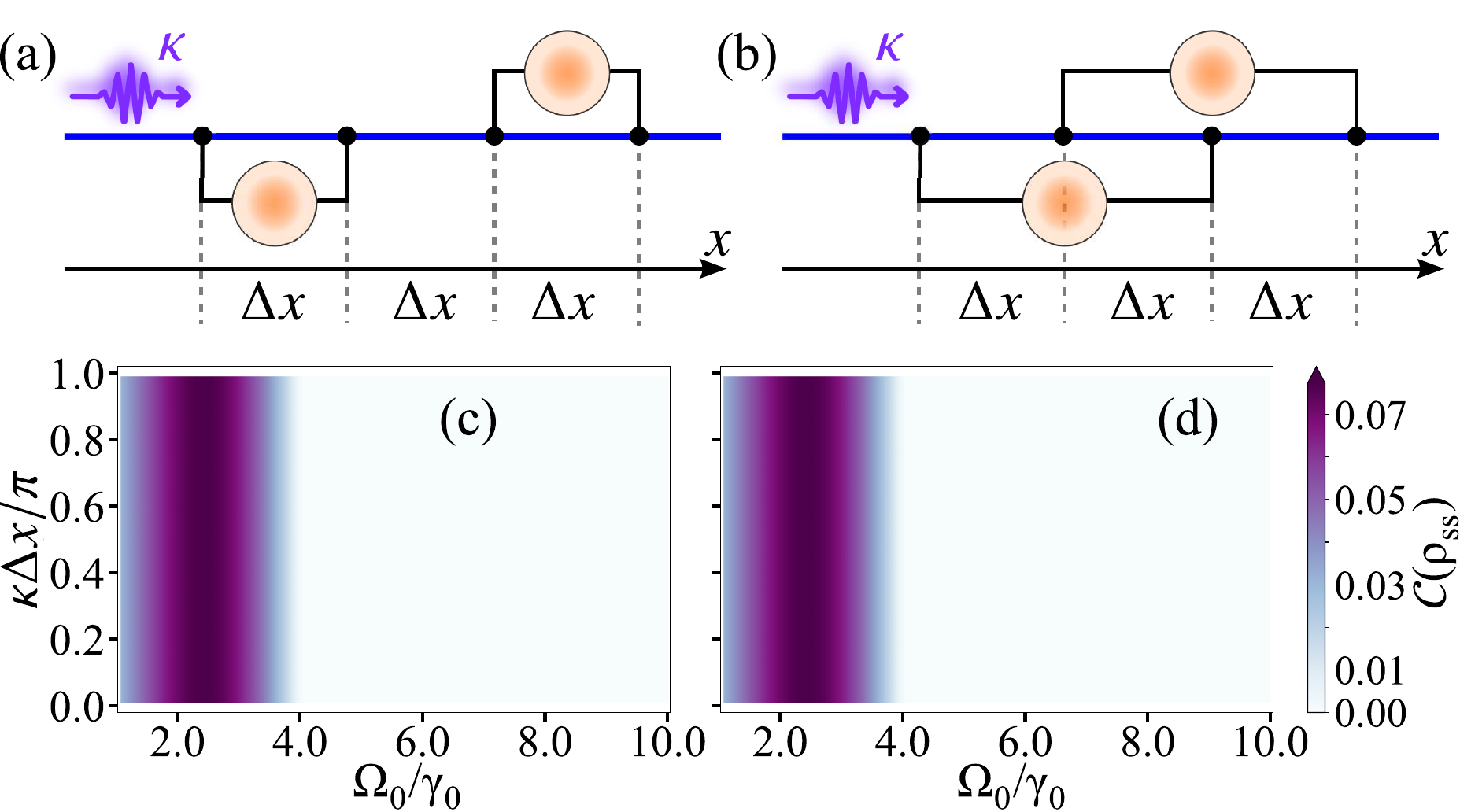}
	\caption{Sketches (a) and (b) show the configuration for giant atoms called Separated and Braided, respectively. From graphs (c) and (d) one sees the amount of entanglement (concurrence) generated in the steady-state for the systems (a) and (b), respectively, as a function of the separation $\kappa \Delta x$ and the external pumping strength $\Omega_{0}$.}
	\label{SupMat-Scheme-concXrabi}
\end{figure}

\section{Steady-state entanglement for other geometries}

For the sake of comparison, in this section we consider two other geometries for giant atoms, namely, Separated and Braided, as shown in Figs.~\ref{SupMat-Scheme-concXrabi}{\color{blue}a} and~\ref{SupMat-Scheme-concXrabi}{\color{blue}b}, respectively. We drive the system using a same approach as in the main text, where an external field is applied to drive the system into the steady-state regime $\rho(t)\rightarrow\rho_{\text{ss}}$. The effective coupling of such an external field with the atom promotes the transition $\ket{gg}\!\rightarrow\!\ket{\psi_{\pm}}$ with Rabi frequency
\begin{align}
	\Omega_{\pm} = \Omega_{0} \left(\frac{\delta_{12} + 2\Delta_{12} \pm \sqrt{4\Delta_{12}^2 + \delta_{12}^2 } }{\sqrt{8\Delta_{12}^2 + 2\delta_{12} \left( \delta_{12} \pm \sqrt{4\Delta_{12}^2 + \delta_{12}^2 } \right)  }}\right) .
\end{align}

The shift $\delta_{12}$ vanishes ($\delta_{1}\!=\!\delta_{2}$) for Braided and Separated atoms, such that we find $\Omega_{-}\!=\!0$ and $\Omega_{+}\!=\!\sqrt{2}\Omega_{0}$. Then, due to the absence of coupling for the transition $\ket{gg}\!\rightarrow\!\ket{\psi_{-}}$, the external field only promotes population inversion from $\ket{gg}$ to the state $\ket{\psi_{+}}$. The decay rate from $\ket{\psi_{\pm}}$ to the state $\ket{gg}$ reads
\begin{align}
	\Gamma_{e\pm}=\Gamma_{\pm g}=\Gamma_{2} \pm \frac{\Gamma_{12}\Delta_{12}}{|\Delta_{12}|} = \Gamma_{2} \pm \Gamma_{12}\mathrm{sgn}(\Delta_{12}) ,
\end{align}
in which we have used that $\Delta_{12}/|\Delta_{12}|=\mathrm{sgn}(\Delta_{12})$. The steady state entanglement cannot be efficiently generated as efficiently as for the nested configuration, since we have $\Gamma_{e\pm}=\Gamma_{\pm g}$, as shown in Figs.~\ref{SupMat-Scheme-concXrabi}{\color{blue}a} and~\ref{SupMat-Scheme-concXrabi}{\color{blue}b}.

\section{Sudden birth of entanglement: More details}

\begin{figure}
	\includegraphics[width=\columnwidth]{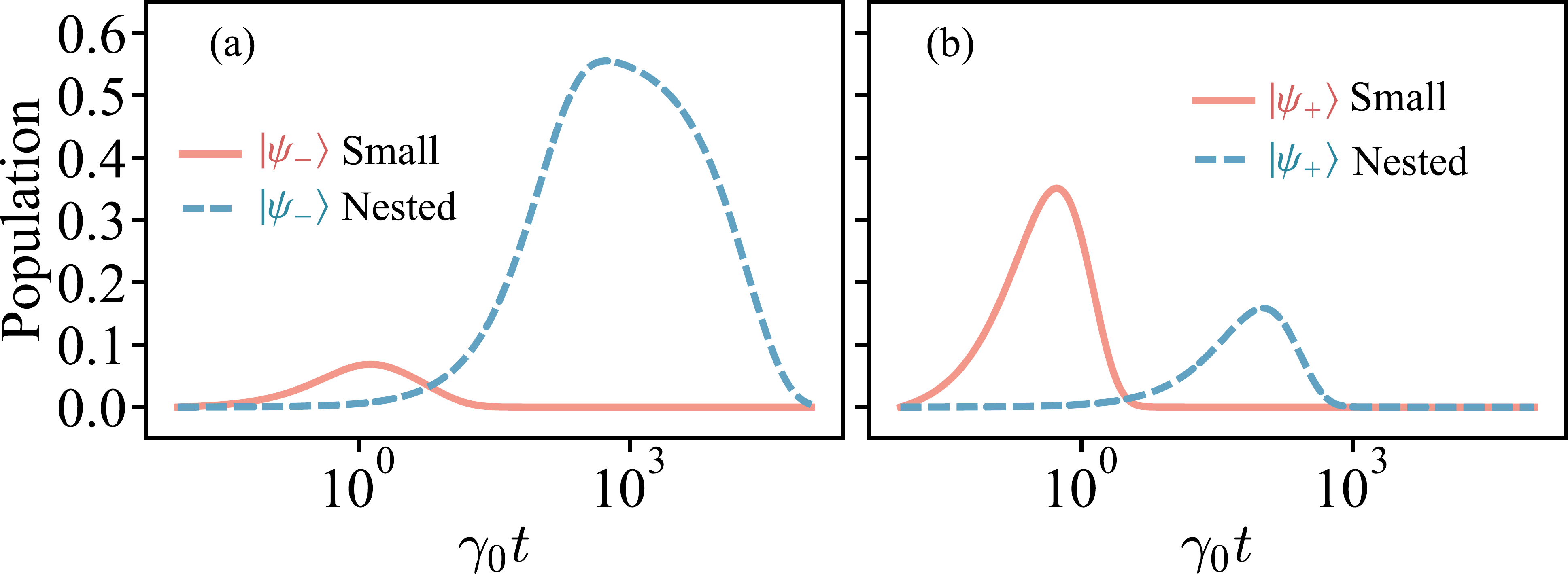}
	\caption{Dynamics of the population in the states $\ket{\psi_{\pm}}$ during the decay dynamics for (a) small and (b) nested giant atoms (giant atoms in nested topology) as a function of $\gamma_{0} t$. We set $\kappa \Delta x \approx 0.19\pi$ for small atoms and $\kappa \Delta x\!=\!0.99\pi$ for giant atoms.}
	\label{SupMat-Populations}
\end{figure}

In this section we present the population dynamics of the modes $\ket{\psi_{\pm}}$ for the situation discussed in the Fig.~4b of the main text. Given the time interval between the maximum of entanglement generation for each system, here we show that it originates in the super- and sub-radiant states in which the system is. In Fig.~\ref{SupMat-Populations} we show the population in each state $\ket{\psi_{\pm}}$, during the decay process, for each geometry considered in the main text. This figure shows that the entanglement created for small atoms and giant atoms in the nested configuration arise in the regime of super- and sub-radiance, respectively. 

\section{Light emitted by giant atoms into a waveguide}

Let us consider the Hamiltonian given by $H = H_{\text{a}} + H_{\text{f}} + H_{\text{c}}$, where $H_{\text{a}}\!=\!\sum_{n=1}^{N}\hbar \omega_{n}\hat{\sigma}_{n}^{+}\hat{\sigma}_{n}^{-}$ is the bare Hamiltonian for the atoms (two-level systems),
\begin{align}
	H_{\text{f}} = \sum_{\nu} \hbar\varpi_{\nu} (a^{\dagger}_{\mathbf{r}\nu}a_{\mathbf{r}\nu} + a^{\dagger}_{\mathbf{l}\nu}a_{\mathbf{l}\nu}) ,
\end{align}
is the field Hamiltonian with creation $a^{\dagger}_{\mathbf{r}\nu}$ ($a^{\dagger}_{\mathbf{l}\nu}$) and annihilation $a_{\mathbf{r}\nu}$ ($a_{\mathbf{l}\nu}$) operators for right-moving (left-moving) bosonic modes with frequency $\varpi_{\nu}$. Finally, the atom-waveguide interaction is described by the coupling Hamiltonian
\begin{align}
	H_{\text{c}} = \hbar \sum_{\nu,n} g_{\nu,n} \left[\hat{\sigma}_{n}^{+}+\hat{\sigma}_{n}^{-} \right] \left[ \alpha^{\ast}_{n\nu} a_{\mathbf{l}\nu} + \alpha_{n\nu} a_{\mathbf{r}\nu} + \mathrm{h.c.}\right] ,
\end{align}
where $g_{\nu,n}$ is coupling strength of the $n$-th atom with $\nu$-th mode, and $\alpha_{n\nu}\!=\!\sum_{j=1}^{K_{n}}e^{-i\varpi_{\nu} x_{j}^{(n)}/c}$ are complex amplitudes that take into account the phase acquired by the bosonic mode at the point $x_{j}^{(n)}$ of the $j$-th connection point of the $n$-th atom with the waveguide.

As a first approximation, let us neglect the terms of the form $\hat{\sigma}_{n}^{-}a_{(\mathbf{l}/\mathbf{r})\nu}$ and $\hat{\sigma}_{n}^{+}a^{\dagger}_{(\mathbf{l}/\mathbf{r})\nu}$, since they can be eliminated by RWA approximation. Then, in the Heisenberg picture we can write the dynamical equation for each mode $a^{\dagger}_{\mathbf{l}\nu}$ and $a^{\dagger}_{\mathbf{r}\nu}$ as
\begin{align}
	\dot{a}^{\dagger}_{(\mathbf{l}/\mathbf{r})\nu} = \frac{i}{\hbar} [H,a^{\dagger}_{(\mathbf{l}/\mathbf{r})\nu}] = \frac{i}{\hbar} [H_{\text{f}},a^{\dagger}_{(\mathbf{l}/\mathbf{r})\nu}] + \frac{i}{\hbar} [H_{\text{c}},a^{\dagger}_{(\mathbf{l}/\mathbf{r})\nu}] .
\end{align}

By using that 
\begin{align}
	[a^{\dagger}_{(\mathbf{l}/\mathbf{r})\nu},a_{(\mathbf{l}/\mathbf{r})\mu}] = \delta_{\nu\mu}\1 , ~~ [a^{\dagger}_{(\mathbf{r}/\mathbf{l})\nu},a_{(\mathbf{l}/\mathbf{r})\mu}] = 0 ,
\end{align}
we obtain
\begin{align}
	[H_{\text{f}},a^{\dagger}_{(\mathbf{l}/\mathbf{r})\nu}] &= \sum_{\mu} \hbar\varpi_{\mu} [(a^{\dagger}_{\mathbf{r}\mu}a_{\mathbf{r}\mu} + a^{\dagger}_{\mathbf{l}\mu}a_{\mathbf{l}\mu}), a^{\dagger}_{(\mathbf{l}/\mathbf{r})\nu} ] \nonumber \\
	&= \hbar\varpi_{\mu} a^{\dagger}_{(\mathbf{l}/\mathbf{r})\nu} .
\end{align}

In addition, the following commutation relation is derived:
\begin{align}
	[H_{\text{c}},a^{\dagger}_{(\mathbf{l}/\mathbf{r})\nu}] &= \hbar \sum_{\mu,n} g_{\mu,n} \hat{\sigma}_{n}^{+} \left[\left( \alpha_{n\mu}^{\ast} a_{\mathbf{l}\mu} + \alpha_{n\mu} a_{\mathbf{r}\mu} \right) , a^{\dagger}_{(\mathbf{l}/\mathbf{r})\nu}\right] \nonumber \\
	&= \hbar \sum_{\mu,n} g_{\mu,n} \hat{\sigma}_{n}^{+} \left( \alpha_{n\mu}^{\ast} \delta_{\mathbf{l}(\mathbf{l}/\mathbf{r})}\delta_{\mu\nu} + \alpha_{n\mu}\delta_{\mathbf{r}(\mathbf{l}/\mathbf{r})}\delta_{\mu\nu} \right) \nonumber \\
	&= \hbar \sum_{n=1}^{N} g_{\nu,n} \hat{\sigma}_{n}^{+} \left( \alpha_{n\nu}^{\ast} \delta_{\mathbf{l}(\mathbf{l}/\mathbf{r})} + \alpha_{n\nu}\delta_{\mathbf{r}(\mathbf{l}/\mathbf{r})} \right) .
\end{align}
where we have used that $\left[a_{\mathbf{l}\mu}, a^{\dagger}_{(\mathbf{l}/\mathbf{r})\nu}\right]\!=\!\delta_{\mathbf{l}(\mathbf{l}/\mathbf{r})}\delta_{\mu\nu}\1$. Therefore, we find
\begin{subequations}
	\begin{align}
		\dot{a}^{\dagger}_{\mathbf{l}\nu} &= i \varpi_{\mu} a^{\dagger}_{\mathbf{l}\nu} + i \sum_{n=1}^{N} \alpha_{n\nu}^{\ast}g_{\nu,n} \hat{\sigma}_{n}^{+} , \\
		\dot{a}^{\dagger}_{\mathbf{r}\nu} &= i \varpi_{\mu} a^{\dagger}_{\mathbf{r}\nu} + i \sum_{n=1}^{N} \alpha_{n\nu}g_{\nu,n} \hat{\sigma}_{n}^{+} .
	\end{align}
\end{subequations}

Now, we can use that
\begin{align}
	\dot{a}^{\dagger}_{\mathbf{l}\nu} - i \varpi_{\mu} a^{\dagger}_{\mathbf{l}\nu} &= e^{i\varpi t} \left[ e^{-i\varpi t} \frac{d a^{\dagger}_{\mathbf{l}\nu}}{dt} - i\varpi e^{-i\varpi t}a^{\dagger}_{\mathbf{l}\nu} \right] \nonumber \\
	&= e^{i\varpi t} \frac{d}{dt}\left(e^{-i\varpi t}a^{\dagger}_{\mathbf{l}\nu}\right) 
\end{align}
to write
\begin{align}
	\frac{d}{dt}\left(e^{-i\varpi_{\nu} t}a^{\dagger}_{\mathbf{l}\nu}\right)  =  i e^{-i\varpi_{\nu} t} \sum_{n=1}^{N} \alpha_{n\nu}^{\ast}g_{\nu,n} \hat{\sigma}_{n}^{+} .
\end{align}

It is then integrated to get
\begin{align}
	a^{\dagger}_{\mathbf{l}\nu}(t) = a^{\dagger}_{\mathbf{l}\nu}(0)e^{i\varpi_{\nu} t} + i \sum_{n=1}^{N} \alpha_{n\nu}^{\ast}g_{\nu,n} \int_{0}^{t}e^{-i\varpi_{\nu} (t^{\prime}-t)} \hat{\sigma}_{n}^{+}(t^{\prime})dt^{\prime} .
\end{align}

In a similar way, we find that
\begin{align}
	a^{\dagger}_{\mathbf{r}\nu}(t) = a^{\dagger}_{\mathbf{r}\nu}(0)e^{i\varpi_{\nu} t} + i \sum_{n=1}^{N} \alpha_{n\nu}g_{\nu,n} \int_{0}^{t}e^{-i\varpi_{\nu} (t^{\prime}-t)} \hat{\sigma}_{n}^{+}(t^{\prime})dt^{\prime} ,
\end{align}
and thus
\begin{subequations}
	\label{a_operators}
	\begin{align}
		a_{\mathbf{l}\nu}(t) &= a_{\mathbf{l}\nu}(0)e^{-i\varpi_{\nu} t} - i \sum_{n=1}^{N} \alpha_{n\nu}g_{\nu,n} \int_{0}^{t}e^{-i\varpi_{\nu} (t-t^{\prime})} \hat{\sigma}_{n}^{-}(t^{\prime})dt^{\prime} ,\\
		a_{\mathbf{r}\nu}(t) &= a_{\mathbf{r}\nu}(0)e^{-i\varpi_{\nu} t} - i \sum_{n=1}^{N} \alpha^{\ast}_{n\nu}g_{\nu,n} \int_{0}^{t}e^{-i\varpi_{\nu} (t-t^{\prime})} \hat{\sigma}_{n}^{-}(t^{\prime})dt^{\prime} .	
	\end{align}
\end{subequations}

Now, by integrating over the frequency $k$, one gets
\begin{align}
	\hat{\Ecal}(t) &= L\int_{-\infty}^{\infty}a_{\mathbf{l}\nu}(t)dk \nonumber \\
	&= \hat{\Ecal}_0(t) - i \sum_{n=1}^{N} g_{n} \int_{0}^{t}L\int_{-\infty}^{\infty}\alpha_{n}(k)e^{ikc (t-t^{\prime})}dk \hat{\sigma}_{n}^{-}(t^{\prime})dt^{\prime} ,
\end{align}
with $\hat{\Ecal}_0(t)\!=\!L\int_{-\infty}^{\infty}a_{\mathbf{l}\nu}(0)e^{-i\varpi_{\nu} t}$, $L$ a normalization parameter. We have here used $\varpi_{\nu}\!=\!-kc$ for the left traveling photon, while the coefficients $\alpha_{n\nu}$ and $g_{n\nu}$ are now functions of $k$ denoted as $\alpha_{n}(k)$ and $g_{n}(k)$, respectively. Now, we assume that the phase-shift at the connection point of the atom varies very slowly with $k$, so that $|\partial_{k}\alpha_{n}(k)|\!\ll\!1$, so we can simplify the above equation as
\begin{align}
	\hat{\Ecal}(t) &= \hat{\Ecal}_{0}(t) - i \sum_{n=1}^{N}\sum_{j=1}^{K_{n}} g_{n} e^{-ik x_{j}^{(n)}}\int_{0}^{t}L\int_{-\infty}^{\infty}e^{ikc (t-t^{\prime})}dk \hat{\sigma}_{n}^{-}(t^{\prime})dt^{\prime} \nonumber \\
	&= \hat{\Ecal}_{0}(t) - i \frac{1}{c}\sum_{n=1}^{N}\sum_{j=1}^{K_{n}} g_{n} e^{-ik x_{j}^{(n)}} \int_{0}^{t} \delta(t-t^{\prime}) \hat{\sigma}_{n}^{-}(t^{\prime})dt^{\prime}
	\nonumber \\
	&= \hat{\Ecal}_{0}(t) - i \frac{1}{c}\sum_{n=1}^{N}\sum_{j=1}^{K_{n}} g_{n} e^{-ik x_{j}^{(n)}}\hat{\sigma}_{n}^{-}(t) .
\end{align}

Then, the above equation shows that the field in the waveguide is a combination of the input field $\hat{E}_{0}(t)\!\propto\!\hat{\Ecal}_{0}(t)$ and the field emitted by the atoms $\hat{E}_{\mathbf{l}}(t)$, which we identify as being the second term in right-hand side. Therefore, we write that the operator of the field emitted by the $N$ atom system is given by
\begin{align}
	\hat{E}_{\mathbf{l}}(t) \propto \sum_{n=1}^{N}\sum_{j=1}^{K_{n}} e^{-ik x_{j}^{(n)}}\hat{\sigma}_{n}^{-}(t) ,
\end{align}
so that the case of small atoms can be recovered by taking $K_{n}\!=\!1$. Now, by considering the operator $a_{\mathbf{r}\nu}(t)$ we can show that
\begin{align}
	\hat{E}_{\mathbf{r}}(t) \propto \sum_{n=1}^{N}\sum_{j=1}^{K_{n}} e^{ik x_{j}^{(n)}}\hat{\sigma}_{n}^{-}(t) .
\end{align}

\section{The $Q$ Mandel parameter}

The $Q$ Mandel parameter was originally proposed as~\cite{Mandel:79,Mandel:Book}
\begin{align}
	Q_{\text{M}} = \frac{\langle\hat{n}^2\rangle - \langle\hat{n}\rangle^2}{\langle\hat{n}\rangle} = \frac{\langle\hat{n}^2\rangle}{\langle\hat{n}\rangle} - \langle\hat{n}\rangle ,
\end{align}
with $\hat n\!=\!a^{\dagger} a$, in which we can use function $g^{(2)}(0)\!=\!\langle\hat{n}^2\rangle/\langle\hat{n}\rangle^2$, so that
\begin{align}
	Q_{\text{M}} = \langle\hat{n}\rangle \left( g^{(2)}(0) - 1 \right) .
\end{align}

Hence, the witness $g^{(2)}`(0)<1$ for quantum light translates into $Q_M<0$ for the Mandel parameter.

\section{Steady state $g^{(2)}(0)$}

By considering the definition of $g^{(2)}(0)$ in the main text, we have
\begin{align}
	g^{(2)}_{\mathbf{l}/\mathbf{r}}(t,t+\tau) = \frac{\langle \hat{E}_{\text{a}}^{\mathbf{l}/\mathbf{r}\dagger}\hat{E}_{\text{a}}^{\mathbf{l}/\mathbf{r}\dagger}\hat{E}_{\text{a}}^{\mathbf{l}/\mathbf{r}}\hat{E}_{\text{a}}^{\mathbf{l}/\mathbf{r}}\rangle_{\rho_{\text{ss}}}} {\langle \hat{E}_{\text{a}}^{\mathbf{l}/\mathbf{r}\dagger}\hat{E}_{\text{a}}^{\mathbf{l}/\mathbf{r}}\rangle_{\rho_{\text{ss}}}^2} ,
\end{align}
where, given a general density matrix $\rho_{\text{ss}}$ for two atoms, we obtain
\begin{align}
	g^{(2)}_{\mathbf{l}/\mathbf{r}}(0) = \frac{4 \varrho_{ee}}{I^2(\kappa \Delta x)} ,
\end{align}
for the nested configuration considered in the paper. Here, we get that
\begin{align}
	I(\kappa \Delta x) &= \langle \hat{E}_{\text{a}}^{\mathbf{l}/\mathbf{r}\dagger}\hat{E}_{\text{a}}^{\mathbf{l}/\mathbf{r}}\rangle_{\rho_{\text{ss}}} = 2\varrho_{ee} + \varrho_{ge} + \varrho_{eg} + 2\varrho_{\text{od}} \cos(2\kappa \Delta x) ,
\end{align}
where $\varrho_{xy}\!=\!\bra{xy}\rho_{\text{ss}}\ket{xy}$, and $\varrho_{\text{od}}\!=\!\bra{eg}\rho_{\text{ss}}\ket{ge}$ are the off-diagonal element of the density matrix relevant for $g^{(2)}_{\mathbf{l}/\mathbf{r}}(0)$. In order to apply this result to the nested case discussed in the main text, we consider the regime $(\kappa \Delta x)^2\!\ll\!1$, so that the above equation rewrites
\begin{align}
	I(\kappa \Delta x)|_{(\kappa \Delta x)^2 \ll 1} &= \langle \hat{E}_{\text{a}}^{\mathbf{l}/\mathbf{r}\dagger}\hat{E}_{\text{a}}^{\mathbf{l}/\mathbf{r}}\rangle_{\rho_{\text{ss}}} = 2\varrho_{ee} + \varrho_{ge} + \varrho_{eg} + 2\varrho_{\text{od}} .
\end{align}

In the case $\kappa \Delta x\!=\!0.01\pi$ we get the numerical solution of $\rho_{\text{ss}}$ where $\varrho_{ee}\!\approx\!0.15\times10^{-4}$, and $\varrho_{ge} + \varrho_{eg} + 2\varrho_{\text{od}}\!\approx\!4.59\times10^{-4}$, and therefore we obtain the giant photon bunching value:
\begin{align}
	g^{(2)}_{\mathbf{l}/\mathbf{r}}(0)|_{\kappa \Delta x=0.01\pi} \approx 2.51 \times 10^2 .
\end{align}

\end{document}